\documentclass{article}
\usepackage{graphicx} % Required for inserting images
\usepackage{hyperref}
\usepackage{subcaption} 
\usepackage{caption} 
 \usepackage{float}
\usepackage{booktabs,tabularx}
\usepackage[margin=1in]{geometry}
\usepackage{amsmath}
\usepackage{todonotes}
\usepackage{authblk}
\usepackage{placeins} % For \FloatBarrier
\usepackage{placeins}
\usepackage{afterpage}
\usepackage{morefloats}
\usepackage{utfsym}

\newcommand{\CM}{\textsc{CheckMATE}}
\newcommand{\rivet}{\textsc{Rivet}}
\newcommand{\contur}{\textsc{Contur}}

\title{Reinterpreting the ATLAS  HHH$\to6b$ Search\\ with \CM{} and \rivet{}:\\ Validation, TRSM Benchmarks, and HL-LHC Prospects}

\author[1]{Tomasz Procter\thanks{\texttt{tomasz.procter@uj.edu.pl}}}
\author[2]{Krzysztof Rolbiecki\thanks{\texttt{krolb@fuw.edu.pl}}}
\author[1]{Andrzej Siódmok\thanks{\texttt{andrzej.siodmok@uj.edu.pl}}}

\affil[1]{Jagiellonian University, ul.~prof.~Stanis\l{}awa \L{}ojasiewicza 11, 30-348 Krak\'{o}w, Poland}
\affil[2]{Faculty of Physics, University of Warsaw, ul.~Pasteura 5, 02-093 Warsaw, Poland}
\date{\today}

\usepackage{amsmath}
\usepackage{amssymb}   % \checkmark
\usepackage{booktabs}  % \toprule etc.
\usepackage{xspace}
\usepackage{siunitx}
\usepackage{tabularx}  % kolumna X
\usepackage{changepage} % adjustwidth
\usepackage[numbers,sort&compress]{natbib}
\newcommand{\pT}{\ensuremath{p_\mathrm{T}}\xspace}

\newcommand{\mHiggs}[1]{\ensuremath{m_{H#1}}\xspace}

\newcommand{\mHHH}{\ensuremath{m_{HHH}}\xspace}

\newcommand{\dAjj}{\ensuremath{\Delta A_{jj}}\xspace}

\newcommand{\RMSmjj}{\ensuremath{\mathrm{RMS}(m_{jj})}\xspace}
\newcommand{\RMSdRjj}{\ensuremath{\mathrm{RMS}(\Delta R_{jj})}\xspace}
\newcommand{\RMSeta}{\ensuremath{\mathrm{RMS}(\eta)}\xspace}
\newcommand{\HTsixj}{\ensuremath{H_\mathrm{T}^{6j}}\xspace}

\newcommand{\dRH}[1]{\ensuremath{\Delta R_{H#1}}\xspace}

\DeclareSIUnit{\GeV}{GeV}

\begin{document}
\newcommand{\htr}[1]{{\color{red} #1}}
\newcommand{\htb}[1]{{\color{blue} #1}}
\newcommand{\htg}[1]{{\color{green} #1}}

{\rightline{OPENMAPP-26-03, MCNET-26-13}}
{\let\newpage\relax\maketitle}
\begin{abstract}
    We present an implementation in \CM{} and \rivet{} of the ATLAS collaboration search for triple Higgs boson production in a six $b$-jets final state. The search relies on event selection using a deep neural network and a statistical model based on the HS3 format. Owing to the rich validation material provided by ATLAS, we perform a thorough validation of neural network input features and exclusion limits in the Standard Model and its extension with two additional singlet scalar fields (TRSM). Finally, we discuss expected performance of the search at the High Luminosity LHC in several scenarios of systematic uncertainty. We present projected exclusion limits for a set of TRSM benchmark models beyond those considered by ATLAS. 
\end{abstract}

\section{Introduction}
Since the discovery of the Higgs boson~\cite{ATLAS:2012yve,CMS:2012qbp}, probing its couplings and the structure of the scalar potential of the Standard Model (SM) has been one of the main goals of experiments at the Large Hadron Collider (LHC)~\cite{ATLAS:2022vkf,CMS:2022dwd,ATLAS-CONF-2025-006,CMS:2026nce}. Although the Higgs triple and quartic self couplings, $\lambda_3$ and $\lambda_4$, are uniquely determined within the SM, they remain elusive for LHC experiments~\cite{Plehn:2005nk,Binoth:2006ym}. Nevertheless, independent measurements of $\lambda_3$ and $\lambda_4$ will be crucial to gain insight into the shape of the Higgs potential~\cite{Stylianou:2023tgg}. On the other hand, a departure from the values predicted by the SM would be an unambiguous signal of the beyond Standard Model (BSM) physics and would have a profound impact on the physics of the early universe~\cite{Horn:2020wif}. 

At the LHC, the main probes of the Higgs triple and quartic couplings are Higgs boson pair production~\cite{ATLAS:2024ish,CMS:2024awa} and Higgs boson triple production~\cite{ATLAS:2024xcs,CMS:2025gos,CMS:2025jkb}. These processes have the potential to constrain the Higgs boson triple and quartic coupling modifiers $\kappa_i = \lambda_i/\lambda_{i}^{\mathrm{SM}}$, where $\lambda_{i}^{\mathrm{SM}}$ denotes the Standard Model value. In particular, searches for Higgs boson pair production at ATLAS and CMS have placed constraints on the triple coupling modifier  $\kappa_3$, yielding 95\% confidence level (CL) intervals of  $-1.35 < \kappa_3 < 6.37$ for CMS~\cite{CMS:2025ngq} and  $-1.6 < \kappa_3 < 7.2$ for ATLAS~\cite{ATLAS:2024ish}. CMS has also probed triple Higgs production in the process $HHH \to b\bar{b}b\bar{b}\gamma\gamma$ (4$b$2$\gamma$) obtaining  $-533 < \kappa_4 < 544$ at 95\% CL~\cite{CMS:2025jkb}, and in the final state $HHH \to b\bar{b}b\bar{b}b\bar{b}$ (6$b$) that resulted in a limit  $-190 < \kappa_4 < 190$ at 95\% CL~\cite{CMS:2025gos}. The final state of 6$b$ has also been investigated by ATLAS, which reported a limit $-240 < \kappa_4 < 240$ at 95\% CL~\cite{ATLAS:2024xcs}. Finally, ATLAS studied prospects for the search for triple Higgs production in the 6$b$ final state at the High Luminosity LHC (HL-LHC)~\cite{ATL-PHYS-PUB-2025-003}. Although the current limits appear to be very weak, results from different processes involving one, two, and three Higgs bosons in the final state provide an opportunity to obtain model independent constraints on the Higgs self couplings at future colliders~\cite{Haisch:2025pql}.

In this paper, we focus on the reinterpretation of the ATLAS $6b$ search, HIGP-2024-32~\cite{ATLAS:2024xcs}.\footnote{At the time of conducting our study, the CMS searches \cite{CMS:2025jkb,CMS:2025gos} were only available as conference notes and did not provide the additional validation materials needed for reinterpretation.} On the technical side, the ATLAS search is very interesting as it utilises advanced experimental techniques, including  both machine learning methods and a statistical model with a simultaneous fit to a histogram of Deep Neural Network (DNN) outputs in the signal and control regions. The implementation is studied in two frameworks widely used for reinterpretation studies: \CM{}~\cite{Dercks:2016npn,Lara:2025cpm,CM_github} and \rivet~\cite{Bierlich:2024vqo}. We perform a detailed comparison between the implementations and the rich set of ATLAS validation material: the ATLAS collaboration provided neural networks in the ONNX format; a statistical model in the HS3~\cite{HS3} format; many validation histograms~\cite{hepdata}; as well as a code snippet in the \textsc{SimpleAnalysis} framework~\cite{simpleanalysiscode}. 

Apart from the SM context, the ATLAS search also looked at constraints on two BSM models: the Two Real Scalar Model (TRSM)~\cite{Robens:2019kga,Papaefstathiou:2020lyp,Robens:2022nnw} and the Dark Matter with CP Violation Model (DM-CPV)~\cite{Chen:2022vac}. In the TRSM, the SM is extended by adding two $\mathbb{Z}_2$-symmetric real singlet scalar fields, $X$ and $S$. In our study we extend the analysis to a set of benchmark models proposed in Ref.~\cite{Karkout:2024ojx}. 

Finally, we discuss the extrapolation of the ATLAS search to the High Luminosity LHC (HL-LHC). We study several scenarios of uncertainty evolution at the future collider, following the ATLAS HL study~\cite{ATL-PHYS-PUB-2025-003}. We also test the set of TRSM benchmarks points and conclude that at least some of them might be accessible for exclusion using the full data set expected at the HL-LHC.  

Multi-Higgs boson production can also arise in other BSM models with extended scalar sectors, e.g.\ in the Non-Minimal Supersymmetric Standard Model (NMSSM)~\cite{Ellwanger:1996gw}, the Next-to-Two Higgs Doublet Model (N2HDM)~\cite{Chen:2013jvg}. These possibilities were considered in Ref.~\cite{Abouabid:2021yvw}, where a number of scenarios with cross sections larger than 10~fb were introduced. Another example is the complex singlet extension of the SM (cxSM)~\cite{Barger:2008jx} studied from the multi-Higgs production perspective in Ref.~\cite{Lane:2024vur}. These models lead to final states including SM Higgs bosons, as well as BSM scalars decaying to $b\bar{b}$. Finally, triple Higgs production can be affected by higher dimension operators~\cite{Li:2026kqk,Roy:2026kud} or anomalous couplings in the effective field theory approach, giving insights into new physics at the multi-TeV scale~\cite{Papaefstathiou:2023uum}. For an extensive review of challenges and opportunities in searches for triple Higgs production, see Ref.~\cite{Abouabid:2024gms}. 

The paper is organised as follows. In Section~\ref{sec:th} we introduce the TRSM used as a benchmark for BSM analysis in this paper. In Section~\ref{sec:atlas} we describe in some detail the ATLAS analysis. In Section~\ref{sec:implementation} we provide the details of implementation of $HHH$ search in \rivet{} and \CM{} and in Section~\ref{sec:validation} the validation of these implementations. Finally, in Sections~\ref{sec:Run2Benchmark} and \ref{sec:hllhc-extrapolation} we study the new set of TRSM benchmark points at $\sqrt{s} = 13$~TeV and in the HL-LHC. We conclude in Section~\ref{sec:conclusion}.

\section{TRSM benchmark models\label{sec:th}}

In this section, we introduce in more detail the TRSM model, which was used by ATLAS to produce training samples for DNNs and for validation of \rivet{} and \CM{} implementations. Finally, we use another set of TRSM benchmark points to demonstrate a physics motivated case of reinterpetation of the triple-Higgs search, in LHC Run-2 and at the HL-LHC.    

In the TRSM, the SM is extended by adding two $\mathbb{Z}_2$-symmetric real singlet scalar fields, $X$ and $S$, resulting in the following scalar potential before the electroweak symmetry breaking~\cite{Robens:2019kga}:
\begin{equation}
\begin{aligned}
        V & = \mu_{\Phi}^2 \Phi^\dagger \Phi + \lambda_{\Phi} {(\Phi^\dagger\Phi)}^2
        + \mu_{S}^2 S^2 + \lambda_S S^4
        + \mu_{X}^2 X^2 + \lambda_X X^4                                              
        + \lambda_{\Phi S} \Phi^\dagger \Phi S^2
        + \lambda_{\Phi X} \Phi^\dagger \Phi X^2
        + \lambda_{SX} S^2 X^2,
\end{aligned}\label{eq:TRSMpot}
\end{equation}
where $\Phi$ is the SM doublet and all couplings are real. There are nine free parameters in the above scalar potential. After electroweak symmetry breaking (EWSB), the additional scalars may obtain vacuum expectation values (vevs), $v_S$ and $v_X$. The gauge eigenstates $\phi_{h,S,X}$ mix through the mixing matrix $R$ to form mass eigenstates $h_{1,2,3}$:
\begin{equation}
    \left( \begin{array}{c} h_1 \\ h_2 \\ h_3 \end{array}\right) =  R \left( \begin{array}{c} \phi_h \\ \phi_S \\ \phi_X \end{array}\right).
\end{equation}
The real matrix $R$ is parametrised by three mixing angles $\theta_{hS}$, $\theta_{hX}$ and $\theta_{SX}$. Throughout this paper we assume for simplicity that the lightest mass eigenstate is the SM-like Higgs boson and fix the following mass ordering: $m_1 < m_2 < m_3$ (see Ref.~\cite{Robens:2019kga} for a general discussion). The nine free parameters of the TRSM can be expressed in terms of the physical masses, mixing angles, and vevs. Fixing the SM Higgs mass to $m_1 = 125$~GeV and its vev $v = 246$~GeV, one is left with seven input parameters that define each benchmark point:
\begin{equation}
    m_2,\quad m_3,\quad \theta_{hS},\quad\theta_{hX},\quad \theta_{SX}, \quad v_S,\quad v_X.\label{eq:TRSMparams}
\end{equation}
The tree-level relations between these parameters and the nine parameters of Eq.~\eqref{eq:TRSMpot} can be found in Ref.~\cite{Robens:2019kga}.

\begin{figure}
    \centering
    \begin{subfigure}[b]{0.25\textwidth}\includegraphics[width=\linewidth]{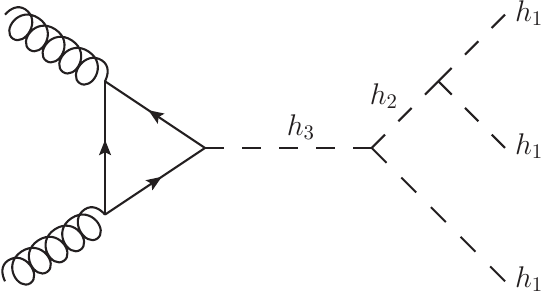}\caption{}\label{fig:feyn(a)}\end{subfigure}\hspace{1cm}
    \begin{subfigure}[b]{0.25\textwidth}\includegraphics[width=\linewidth]{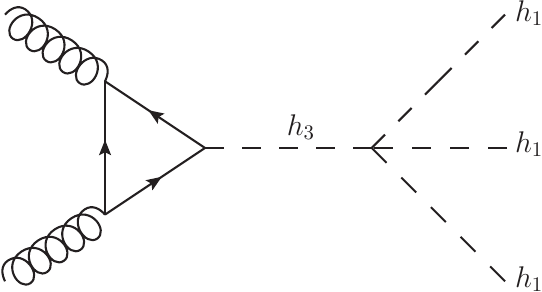}\caption{}\label{fig:feyn(b)}\end{subfigure}\hspace{1cm}
    \begin{subfigure}[b]{0.22\textwidth}\includegraphics[width=\linewidth]{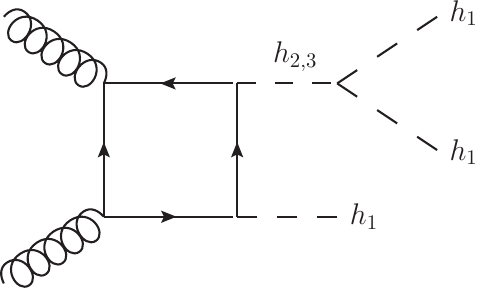}\caption{}\label{fig:feyn(c)}\end{subfigure}
    \caption{Leading order triple Higgs boson production in the TRSM model.}
    \label{fig:resonantTRSM}
\end{figure}

Triple-Higgs production occurs in the TRSM at the leading order (LO) in $pp$ collisions through the single and double-resonant processes shown in Figure~\ref{fig:resonantTRSM}.  Depending on the coupling structure, e.g.\ the size of triple and quartic couplings, different diagrams can dominate the process.
  
Reference~\cite{Robens:2019kga} defined ``Benchmark Plane 3" (BP3), which fulfills theoretical (correct scalar potential minimum) and experimental (Higgs physics) constraints, while maximizing triple Higgs boson production. In this setup, the masses of heavy scalars are allowed to vary in the following ranges $m_2 \in [125, 500]$~GeV, $m_3 \in [255, 650]$~GeV, while the remaining parameters are fixed to: 
\begin{equation}
    \theta_{hS} = -0.129,\quad\theta_{hX}=0.226,\quad \theta_{SX}=-0.899, \quad v_S = 140\ \mathrm{GeV},\quad v_X = 100\ \mathrm{GeV}.
\end{equation}
For BP3 the parameter $\kappa$ that rescales the couplings of the lightest SM-like scalar, relative to the SM couplings, is given by $\kappa = 0.966$. The points from BP3 were used by ATLAS for training of the resonant DNN and for validation in this paper. 

Another set of training points, the heavy-resonant model, is based on TRSM BP3, but with the masses of scalars treated as free parameters. The usual relations between TRSM parameters do not hold anymore, but this approach can be understood as a ``simplified model" for $HHH$ production at the LHC, see also Ref.~\cite{Papaefstathiou:2025meh}. Additionally, only the double-resonant contribution is considered in this case, see Figure~\ref{fig:feyn(a)}. The masses are larger than in the TRSM BP3 set: $600 < m_2 < m_3 < 1500$~GeV. The widths of the scalars are set to 1\% (20\%) that define narrow (wide) width signal models. 

The BP3 points suffer from perturbativity issues due to the large quartic coupling already close to the electroweak scale. Therefore, in Ref.~\cite{Karkout:2024ojx} a new set of benchmark points was constructed, which obey the perturbativity constraint at the EW scale and provide an enhanced cross section, at least 100 times, with respect to the SM prediction for $HHH$ production. The points originate from a scan of a seven-dimensional parameter space, see Eq.~\eqref{eq:TRSMparams}, where 530,000 points were sampled. They generally fall into a doubly resonant category: 
\begin{equation}
pp \to h_3 \to h_2 (\to h_1 h_1)h_1    ,
\end{equation}
shown in Figure~\ref{fig:feyn(a)}. 
These points, which we will describe as the \textit{Optimised Double-Resonant Benchmark} (ODRB) points, to help distinguish from BP3, form an interesting case for a reinterpretation study. We consider them in this work for: testing the sensitivity of the ATLAS $HHH$ search at $\sqrt{s} = 13$~TeV in Section~\ref{sec:Run2Benchmark}; and checking for constraints using a projection to the HL-LHC in different scenarios for systematic errors in the future experiment in Section~\ref{sec:hllhc-extrapolation}.

\section{ATLAS HIGP-2024-32 analysis \label{sec:atlas}}

This work concentrates on the recent ATLAS search for triple Higgs boson production in the six-\(b\)-jet final state, based on \(pp\) collision data at \(\sqrt{s}=13~\text{TeV}\) with an integrated luminosity of \(126~\text{fb}^{-1}\) collected during Run~2 of the LHC~\cite{ATLAS:2024xcs}. The analysis targets scenarios where three Higgs bosons are produced and each decays as \(H\to b\bar b\), leading to a final state with six \(b\)-jets. The results are interpreted for both non-resonant and resonant production modes, including SM-like triple-Higgs production and BSM benchmark models featuring additional scalar resonances, namely the TRSM and the DM-CPV.

In the analysis, hadronic jets are reconstructed from particle-flow objects using the anti-\(k_t\) algorithm~\cite{Cacciari:2008gp} with radius parameter \(R=0.4\). Analysis event selection requires at least six reconstructed jets with \(p_{\mathrm{T}}>20~\text{GeV}\) and \(|\eta|<2.5\), with the leading four jets having \(p_{\mathrm{T}}>40~\text{GeV}\).  Bottom-quark initiated jets are identified using the DL1d deep neural network \(b\)-tagging algorithm~\cite{ATLAS:2022gqe,ATLAS:2020jip,ATLAS:2022qxm,ATLAS:2025dkv}. In this paper we will focus on describing the 6\(b\) selection which is used as the signal region, though we also simulate the signal contribution to the 5$b$ control region for use in the statistical fit.

The six \(b\)-tagged jets are paired to form three Higgs boson candidates. The pairing is chosen by minimizing
\begin{equation}
\left|m_{H_1}-120~\text{GeV}\right|
+ \left|m_{H_2}-115~\text{GeV}\right|
+ \left|m_{H_3}-110~\text{GeV}\right|,
\label{equation:higgs_candidate_pairing_cost_function}
\end{equation}
over all possible assignments of the six jets into three disjoint pairs, subject to the transverse-momentum ordering constraint
\begin{equation}
p_{\mathrm{T}}^{ab} > p_{\mathrm{T}}^{cd} > p_{\mathrm{T}}^{ef},
\end{equation}
where the pairs \((a,b)\), \((c,d)\), and \((e,f)\) are associated with the leading, subleading, and third Higgs candidates \(H_1\), \(H_2\), and \(H_3\), respectively. The reference masses \((120, 115, 110)~\text{GeV}\) are chosen to follow the peaks of the reconstructed di-jet mass distributions in simulated signal samples, accounting for detector effects and energy losses.

The final discrimination between signal and background is provided by DNN classifiers trained separately for three kinematic regimes: non-resonant, resonant, and heavy-resonant triple-Higgs production. For the resonant and non-resonant regimes, the mass ranges considered are $325 < m_3 < 575$~GeV and $200 < m_2 < 350$~GeV. Additionally, in the resonant category we have $m_3 > m_2 + m_H$ and $m_2 > 2m_H$. For the heavy-resonant regime the mass ranges are $550 \leq m_3 \leq 1500$~GeV and $275 \leq m_2 \leq 1000$~GeV. The three DNNs, denoted \texttt{nonresDNN}, \texttt{resDNN}, and \texttt{heavyresDNN}, are trained using data from the 5$b$ region as a common background-dominated sample, together with simulated signal samples that differ by regime: the \texttt{nonresDNN} uses SM $HHH$ production, while the \texttt{resDNN} and \texttt{heavyresDNN} each use an equally weighted mixture of all the relevant signal samples spanning their respective mass ranges. The resonant classifiers are therefore optimised for broad sensitivity across the mass range rather than for any individual resonance point. As summarised in Table 1, the input variables are optimised separately for each network, with ten kinematic variables chosen from a larger set based on their separation power and their limited correlation with the $b$-jet multiplicity, in order to keep the background extrapolation uncertainties under control.

% The input variables include combinations of reconstructed Higgs candidate masses, global event-shape observables (aplanarity, sphericity, transverse sphericity), angular separations between jets, and scalar sums of jet transverse momenta, see Table~\ref{tab:inputDNN}. 

\begin{table}
\begin{minipage}{\textwidth}
\centering
\begin{adjustwidth}{-0.0cm}{0.0cm} % <- tabela 1.6 cm szersza niż tekst (0.8 cm na stronę)
\caption{Summary of the input variables used in each DNN.}
\label{tab:inputDNN}
\begin{tabularx}{\dimexpr\textwidth\relax}{p{2.4cm}Xp{1cm}p{0.6cm}p{1cm}}
\toprule
Variable & Definition & nonres & res & heavyres\\
\midrule
\({m_H\mbox{-}\mathrm{radius}}\) &
Distance between the event (\(\mHiggs{1}, \mHiggs{2}, \mHiggs{3}\)) vector and expected vector for signal events,
\((120, 115, 110)~\SI{}{\GeV}\). &  \usym{1F5F8} & &  \usym{1F5F8} \\
\midrule
\mHiggs{1} &
Reconstructed mass of the highest \pT\ Higgs boson candidate. &\usym{1F5F8}& &\usym{1F5F8} \\
\midrule
\RMSmjj &
Root-mean-squared (RMS)\footnote{One point which is clear from the \textsc{SimpleAnalysis} code provided, but nevertheless may surprise readers unfamiliar with ROOT nomenclature, is that the term ``root-mean-square'', as used in the paper, actually refers to the quantity more commonly known as the standard deviation. This is a legacy of the function names in ROOT~\cite{Brun:1997pa}.}. of the invariant mass of all 15 possible jet pairs. &\usym{1F5F8}& &\usym{1F5F8} \\
\midrule
\RMSdRjj &
RMS of the angular separation between all 15 possible jet pairs. & \usym{1F5F8}& \usym{1F5F8} &\usym{1F5F8} \\
\midrule
\RMSeta &
RMS of the pseudo-rapidity of the six candidate jets, before pairing into Higgs boson candidates\footnote{As will be discussed in Section~\ref{subsec:NN_input_validation}, this was missed in Table~1 of Reference~\cite{ATLAS:2024xcs}.}. & \usym{1F5F8}& &\usym{1F5F8} \\
\midrule
Skewness \dAjj &
Skewness of \(\cosh(\Delta \eta_{ik}) - \cos(\Delta \phi_{ik})\), where \(i, k\) are all 15 possible jet pairs. & &\usym{1F5F8}& \\
\midrule
\HTsixj &
Scalar sum of the \pT\ of the 6 jets selected to reconstruct the 3 Higgs boson candidates. & &\usym{1F5F8} & \\
\midrule
\(\cos\theta\) &
In the \((\mHiggs{1}, \mHiggs{2}, \mHiggs{3})\) coordinate system, \(\theta\) is the angle between the vector from the origin
to the event's reconstructed mass of the Higgs boson candidates, and the vector from the origin to
\((120, 115, 110)~\SI{}{\GeV}\). & &\usym{1F5F8} & \\
\midrule
Aplanarity\(_{6\text{j}}\) &
The fraction of \pT\ from the 6 jets selected lying outside the plane formed by the 2 highest \pT\ jets \cite{ATLAS:2012tch}. &\usym{1F5F8} &\usym{1F5F8} & \usym{1F5F8} \\
\midrule
Sphericity\(_{6\text{j}}\) &
Isotropy of the momenta of the 6 jets selected to reconstruct the 3 Higgs boson candidates \cite{ATLAS:2012tch}. & &\usym{1F5F8}  & \\
\midrule
Transverse Sphericity\(_{6\text{j}}\) &
Isotropy of the \pT\ of the 6 jets used for Higgs reconstruction, within the \(x-y\) plane \cite{ATLAS:2012tch}. &\usym{1F5F8} & & \\
\midrule
Sphericity &
Isotropy of the momenta of all jets in the event \cite{ATLAS:2012tch}.& & & \usym{1F5F8}  \\
\midrule
\mbox{\(\eta-\mHHH\)} \mbox{fraction} &
\(
\frac{\sum_{i,k} 2 \pT^{i} \cdot \pT^{k} \cdot (\cosh(\Delta \eta(ik)) - 1)}{\mHHH^2}
\)
where \(i, k\) are all 15 possible jet pairs, and \mHHH\ is the reconstructed tri-Higgs invariant mass. & &\usym{1F5F8} & \\
\midrule
\dRH{1} &
Angular separation between the jets paired to form the highest \pT\ Higgs boson candidate. &\usym{1F5F8} &\usym{1F5F8} &\usym{1F5F8} \\
\midrule
\(\Delta R_{H2}\) &
Angular separation between the jets paired to form the second-highest \pT\ Higgs boson candidate. &\usym{1F5F8} &\usym{1F5F8} &\usym{1F5F8} \\
\midrule
\(\Delta R_{H3}\) &
Angular separation between the jets paired to form the lowest \pT\ Higgs boson candidate.&\usym{1F5F8} &\usym{1F5F8} &\usym{1F5F8} \\
\bottomrule
\end{tabularx}

\end{adjustwidth}
\label{tab:strategy:vars}
\end{minipage}
\end{table}

%The DNNs are implemented with fully connected architectures using rectified linear unit activations and are trained to output a score between 0 and 1, with signal events populating the high-score region.

\section{Implementation in \rivet{} and \CM\label{sec:implementation}}

\subsection{Event generation for validation}
\label{subsec:implementation_evgen}

SM and TRSM (BP3 and heavy resonance) signal events at $\sqrt{s} = 13$~TeV and $\sqrt{s} = 14$~TeV were generated using \textsc{MadGraph5\_aMC@NLO 3.5.5}~\cite{Alwall:2014hca,Alwall:2007fs,Alwall:2008qv,Hirschi:2015iia} with the NNPDF23LO~\cite{Ball:2012cx,Buckley:2014ana,NNPDF:2014otw} parton distribution function (PDF) set. The events were then interfaced with \textsc{Pythia}~8.3~\cite{Sjostrand:2014zea,Bierlich:2022pfr} to model decays, hadronization, and showering. The number of events for each validation point was typically in the range of 200k--250k. 

The SM $HHH$ production cross section at the next-to-next-to-leading-order (NNLO) in the gluon-fusion process is taken as $\sigma^{\textrm{SM}}_{HHH} = 0.079^{+0.012}_{-0.013}$~fb~\cite{deFlorian:2019app} at the center-of-mass energy 13~TeV, and $\sigma^{\textrm{SM}}_{HHH} = 0.103^{+5\%}_{-8\%}$~fb~\cite{deFlorian:2019app} at 14~TeV (implying a scale factor 1.30). The $H\to b\bar{b}$ branching ratio (BR) is set to $0.582$ following the recommendations of~\cite{LHCHiggsCrossSectionWorkingGroup:2016ypw} for the Higgs boson mass of 125~GeV.

The parameter cards are generated using the TRSM repository~\cite{gitlabrepoTRSM} which also provides UFO~\cite{Darme:2023jdn} implementation of the TRSM for  \textsc{MadGraph5\_aMC@NLO}. The decay tables  and widths were automatically recalculated in \textsc{MadGraph} for consistency~\cite{Alwall:2014bza}, but decays are simulated in \textsc{Pythia}.\footnote{The event generation command is therefore: \texttt{generate p p > h h h [noborn=QCD]}.} The TRSM benchmarks include 27 points parametrised by the masses of heavy scalars $(m_S,m_X) \equiv (m_2, m_3)$.\footnote{In the ATLAS notation, the lighter of the scalars is labeled $S$ while the heavier $X$. This is in conflict with the notation introduced in Section~\ref{sec:th}, where these labels denoted gauge singlet fields. } The points are categorised as ``resonant" (19 points), for which $m_X > m_S + m_h$ and $m_S > 2 m_h$ and the resonant cascade decay is dominant, and ``non-resonant" (8 points) for which non-resonant diagrams are dominant.\footnote{Note that the validation material (HS3 workspaces) does not automatically follow this convention, and e.g.\ for the model point $(275,550)$ the non-resonant statistical model is published. Interestingly the limit from the non-resonant DNN is stronger than the resonant one, even though the point is clearly double-resonant. } Finally a class of 45 points in the heavy-resonant category is considered, which represent a simplified phenomenology with cascade resonant production and fixed decay widths of 1\% and 20\% of the scalars $X$ and $S$. The cross section for BSM models in the MC chain is fixed at 50~fb (many of the TRSM benchmark points in fact have the LO cross section in the range 20--30~fb). 

\subsection{CheckMATE}
\label{subsec:implementation_cm}

Implementation in \CM{}~\cite{Dercks:2016npn,Lara:2025cpm} follows a standard workflow, efficiency functions, and smearing parameters as for other implemented 13 TeV ATLAS analyses. The only difference is in the $b$-jet tagging efficiency function. Following the ATLAS search, the implementation in \CM{} uses the \texttt{DL1d} algorithm digitised from Figure~1a of~\cite{dl1d}. The implementation includes signal and control regions. For DNN inference, the ONNX Runtime C++ library is used~\cite{onnxruntime}. Simulated events are subject to a fast detector simulation performed using \textsc{Delphes~3.5}~\cite{deFavereau:2013fsa}. Jets are reconstructed using \textsc{FastJet}~\cite{Cacciari:2011ma} and the anti-$k_t$ algorithm~\cite{Cacciari:2008gp}.  

For statistical evaluation, a new interface for HS3-format likelihoods was developed. Original likelihoods were changed by removing experiment-specific signal modifiers. The luminosity constraint and statistical uncertainty were included in the final model. The evaluation of the statistical model is performed using \texttt{XRooFit} from the \texttt{ROOT.Experimental} module.

\subsection{Rivet}
\label{subsec:implementation_rivet}

The analysis was also implemented in the \rivet{} \cite{Bierlich:2024vqo} framework. The existing library of smearing functions designed for ATLAS Run~2 was used for detector emulation (as with other detector-level analyses contained in \rivet{}), with extensions to the $b$-tagging functions (DL1d) added specifically for this analysis. 

Typically, statistical inference using the results of \rivet{} analyses is carried out using the \contur{}~\cite{CONTUR:2025yis} package. However, in this case, this was not practical: \contur{} does not interface with ROOT, and (for the moment) the HS3 likelihoods provided by the experiment can only be used for inference with ROOT. We hope that future developments in \textsc{Spey}~\cite{Araz:2023bwx} (to which \contur{} already has an experimental interface) and pyHS3~\cite{pyhs3} will resolve this: in the meantime, statistical computations were carried out by a standalone python script which reads in the necessary yoda and HS3 files, and performs calculations using \texttt{XRooFit},\footnote{An example script is contained in the Zenodo repository~\cite{siodmok_2026_19735119} associated with this publication.} with the likelihoods adjusted as for \CM{} above.

%\section{Results}

\section{Validation}
\label{sec:validation}
We believe this analysis was the first to share validation plots for all DNN inputs in the signal region (as opposed to, for example, just showing a single ``most-important'' feature); and indeed does so for multiple different parameter points. As will be highlighted in the following, these plots were absolutely essential in identifying the exact definitions of some of the input variables, helping to justify why the provision of such plots was one of the primary requests of the Les Houches guide to reinterpretable ML~\cite{Araz:2023mda}. The HS3 files provided on HEPData also allowed a detailed validation of the DNN output distributions and of the statistical procedure used to extract limits from the analysis.

\begin{figure}[!htbp]
    \centering    
    \begin{subfigure}{0.4\textwidth}
        \centering
        \includegraphics[width=\linewidth]{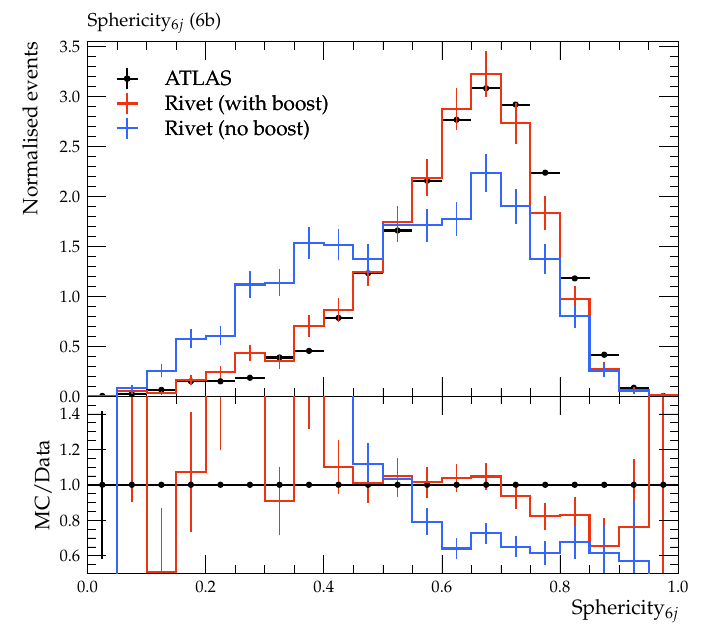}
        \caption{Sphericity (6 jets)}
        \label{subfig:BoostImpactSphericity}
    \end{subfigure}
    \begin{subfigure}{0.4\textwidth}
        \centering
        \includegraphics[width=\linewidth]{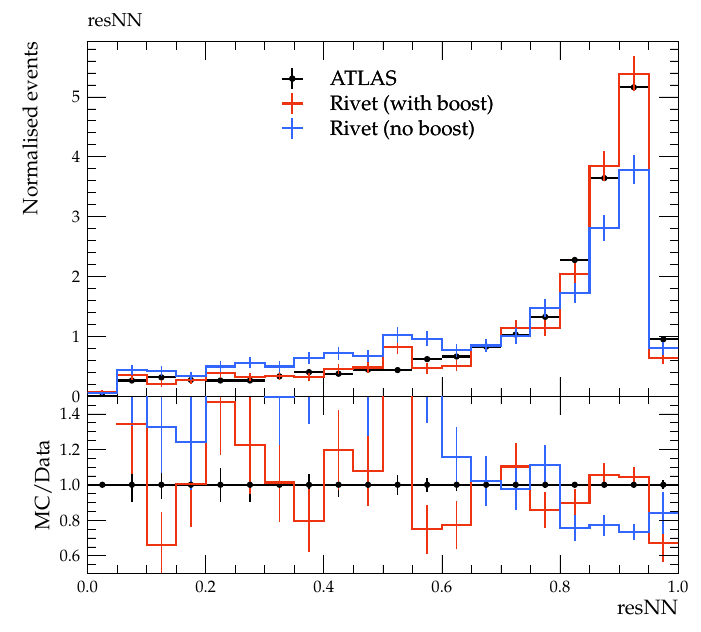}
        \caption{Resonant DNN score}
        \label{subfig:BoostImpactResNN}
    \end{subfigure}
    \caption{The impact of boosting the six-leading jets into the rest frame of the $HHH$ system for calculating the sphericity and associated variables, on a) the sphericity defined in Table~\ref{tab:inputDNN}, and b) the final DNN score, using the (325, 520) resonant benchmark point.\label{fig:BoostImpact}}
\end{figure}

\subsection{Neural network inputs}
\label{subsec:NN_input_validation}

\begin{figure}[!htbp]
    \centering    
    % --- First row (3 images) ---
    \begin{subfigure}{0.27\textwidth}
        \centering
        \includegraphics[width=\linewidth]{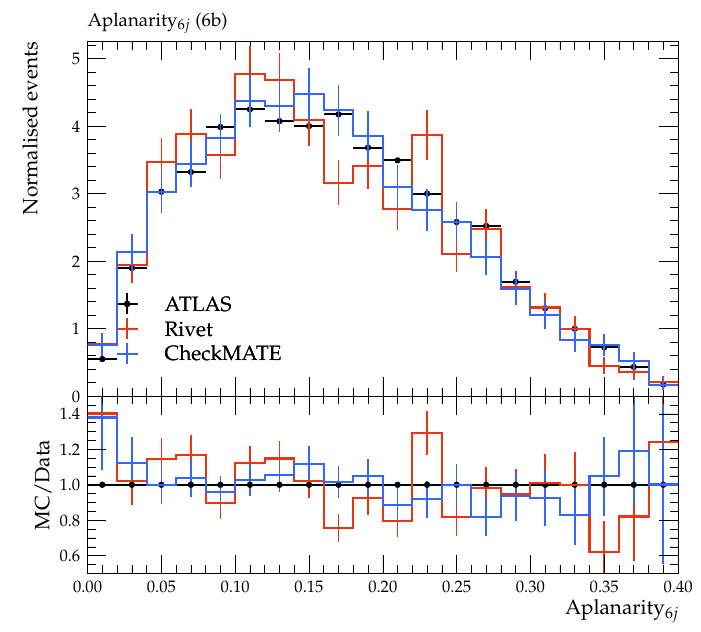}
        \caption{Aplanarity (6 jets)}
        \label{subfig:validation450-275_aplanarity6j}
    \end{subfigure}
    \begin{subfigure}{0.27\textwidth}
        \centering
        \includegraphics[width=\linewidth]{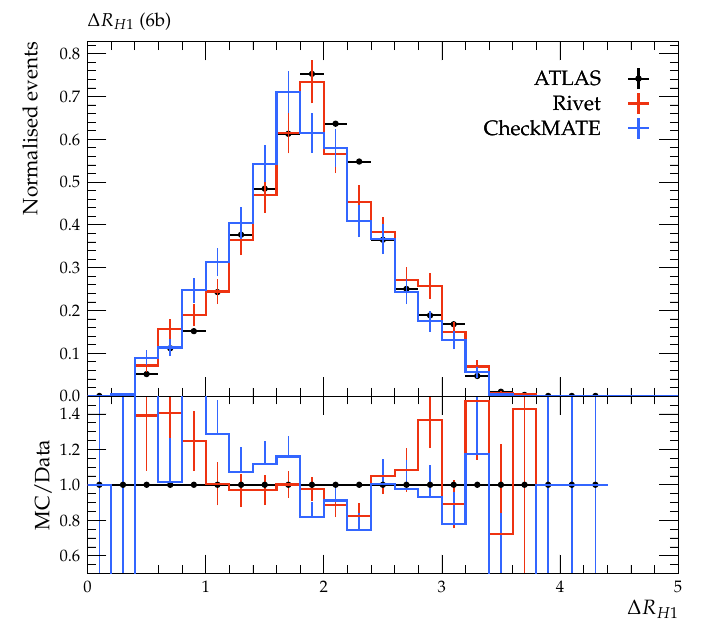}
        \caption{$\Delta R_{H1}$}
        \label{subfig:validation450-275_DeltaRH1}
    \end{subfigure}
    \begin{subfigure}{0.27\textwidth}
        \centering
        \includegraphics[width=\linewidth]{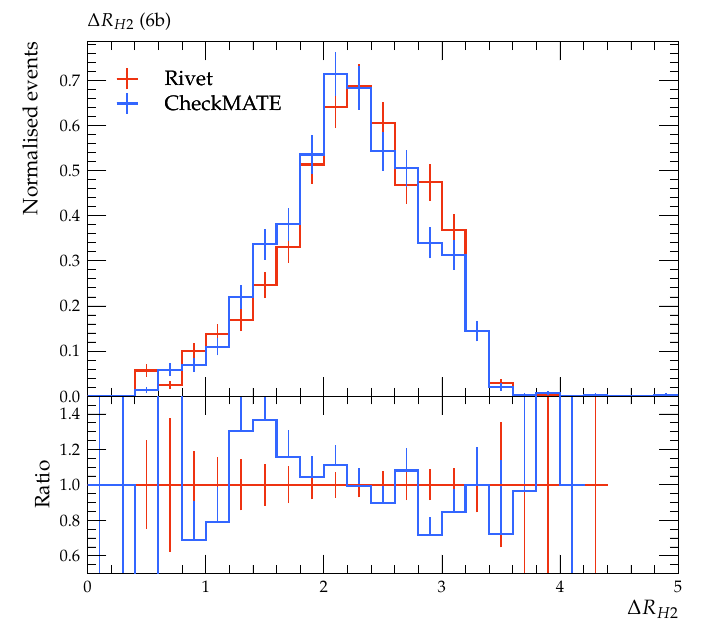}
        \caption{$\Delta R_{H2}$}
        \label{subfig:validation450-275_DeltaRH2}
    \end{subfigure}
    
    \vspace{1em}
    
    % --- Second row (3 images) ---
    \begin{subfigure}{0.27\textwidth}
        \centering
        \includegraphics[width=\linewidth]{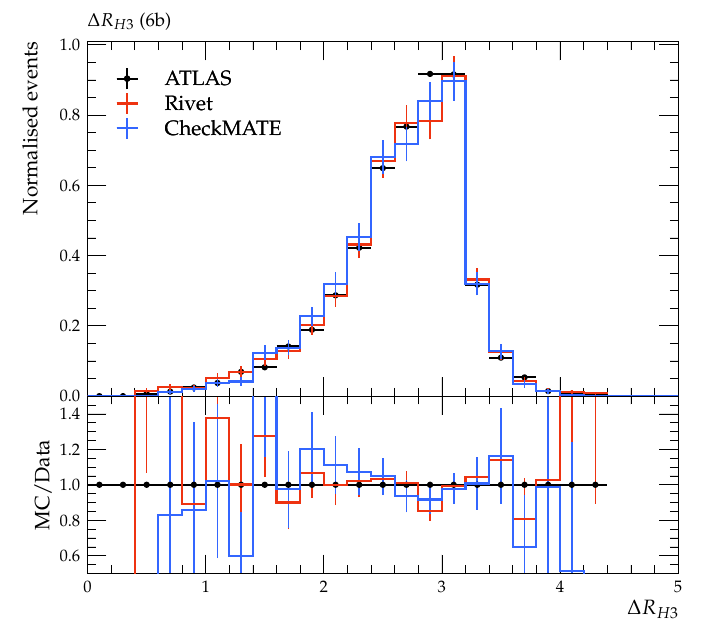}
        \caption{$\Delta R_{H3}$}
        \label{subfig:validation450-275_DeltaRH3}
    \end{subfigure}
    \begin{subfigure}{0.27\textwidth}
        \centering
        \includegraphics[width=\linewidth]{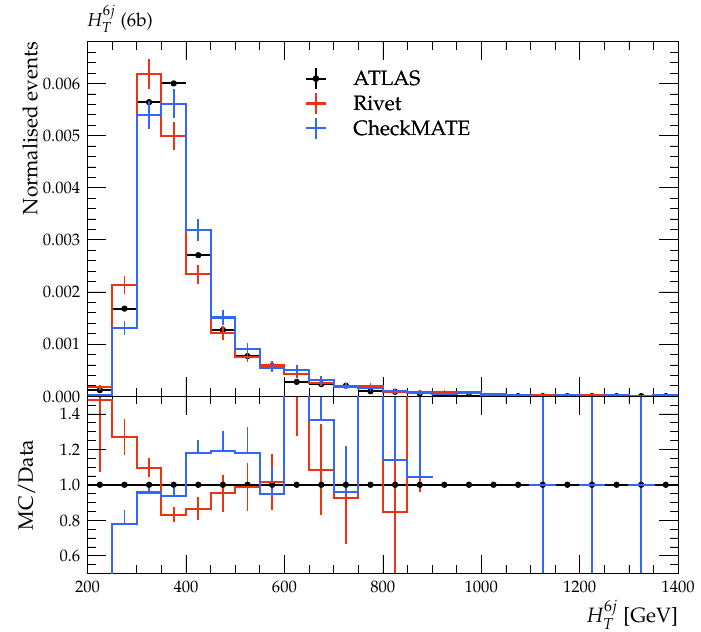}
        \caption{$H_{T}^{6j}$}
        \label{subfig:validation450-275_HT6j}
    \end{subfigure}
    \begin{subfigure}{0.27\textwidth}
        \centering
        \includegraphics[width=\linewidth]{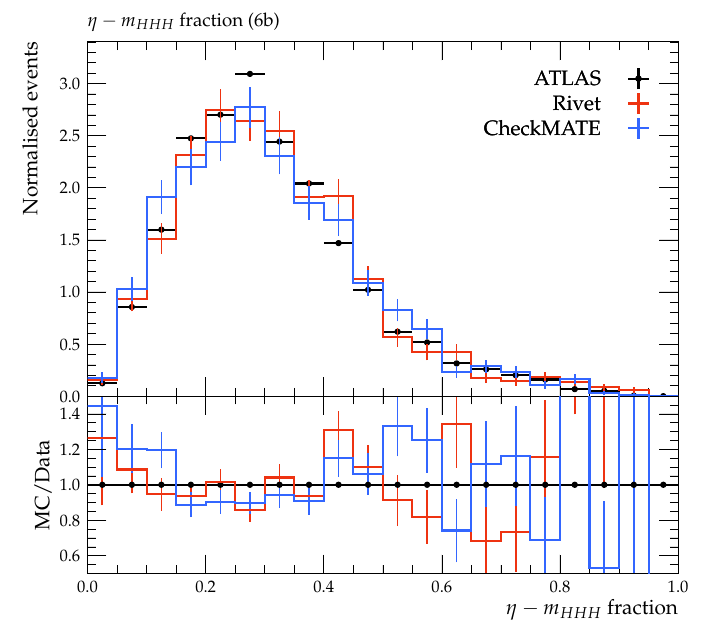}
        \caption{$\eta-m_{HHH}$ fraction}
        \label{subfig:validation450-275_eta_mHHH}
    \end{subfigure}
    
    \vspace{1em}
    
    % --- Third row (3 images) ---
    \begin{subfigure}{0.27\textwidth}
        \centering
        \includegraphics[width=\linewidth]{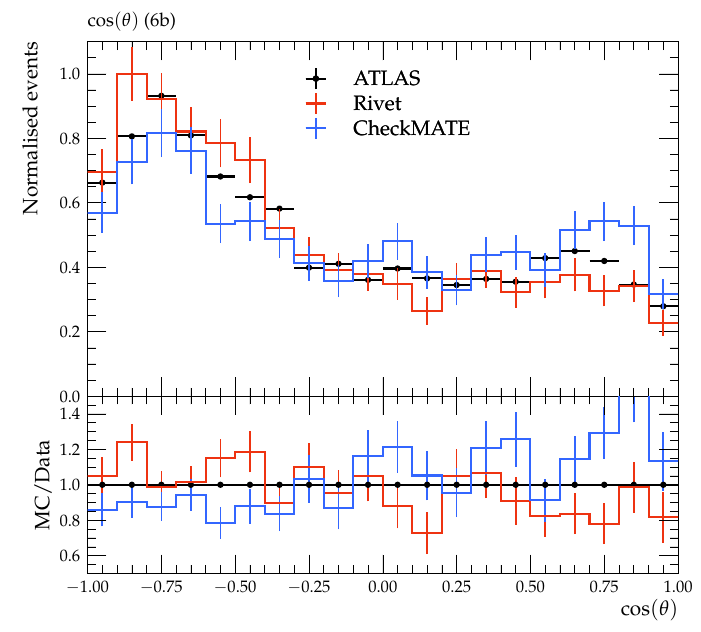}
        \caption{$\cos (\theta)$}
        \label{subfig:validation450-275_CosTheta}
    \end{subfigure}
    \begin{subfigure}{0.27\textwidth}
        \centering
        \includegraphics[width=\linewidth]{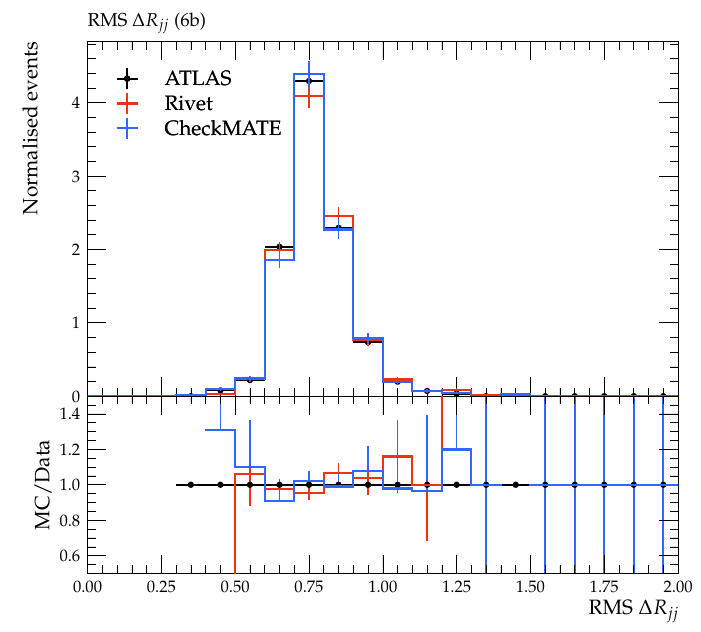}
        \caption{RMS ($\Delta R_{jj}$) }
        \label{subfig:validation450-275_RMSeta}
    \end{subfigure}
    \begin{subfigure}{0.27\textwidth}
        \centering
        \includegraphics[width=\linewidth]{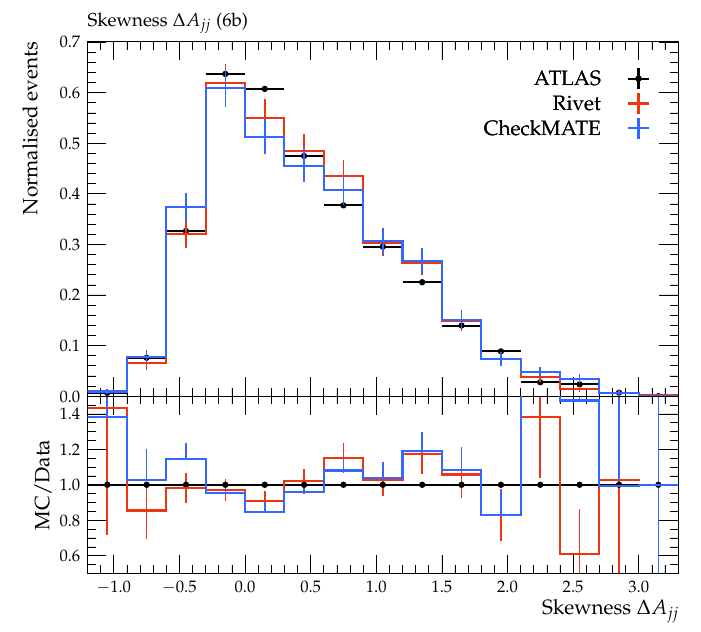}
        \caption{Skewness ($\Delta A_{jj}$) }
        \label{subfig:validation450-275_RMSmjj}
    \end{subfigure}
    
    \vspace{1em}
    
    % --- Last row (2 centered images) ---
    \begin{subfigure}{0.27\textwidth}
        \centering
        \includegraphics[width=\linewidth]{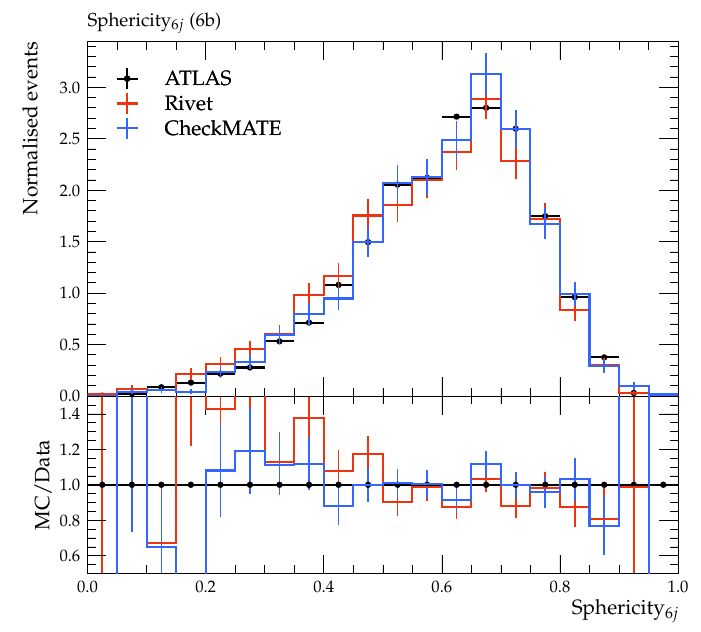}
        \caption{Sphericity (6 jets)}
    \end{subfigure}\hspace{3em} % space between the two
    \begin{subfigure}{0.27\textwidth}
        \centering
        \includegraphics[width=\linewidth]{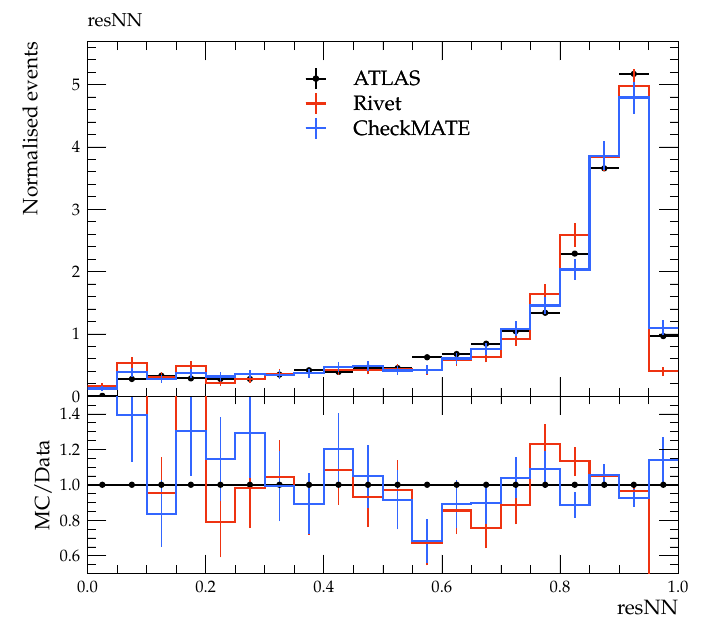}
        \caption{Resonant NN score}
        \label{subfig:validation450-275_NN}
    \end{subfigure}
    
    \caption{The ten normalised NN inputs (a--j) and the normalised NN output (k) for the resonant model, for the (275, 450) mass point, comparing to the results provided by ATLAS (auxiliary figures 5a--7b for the inputs and 17b for the output~\cite{HIGP-2024-32}). \rivet{} was run on 250k events and \CM{} on 200k. Both implementations are in satisfactory agreement, with differences within 20\% in all relevant bins (i.e., those with non-negligible counts) for inputs.}
    \label{fig:TRSM_450_275_validation}
\end{figure}

\begin{figure}[!htbp]
    \centering    
    % --- First row (3 images) ---
    \begin{subfigure}{0.27\textwidth}
        \centering
        \includegraphics[width=\linewidth]{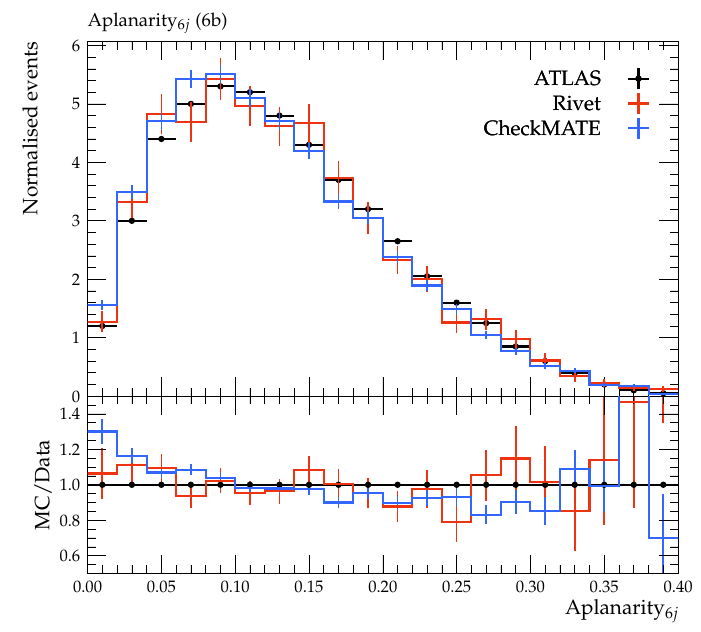}
        \caption{Aplanarity (6 jets)}
        \label{subfig:validationSM_aplanarity6j}
    \end{subfigure}
    \begin{subfigure}{0.27\textwidth}
        \centering
        \includegraphics[width=\linewidth]{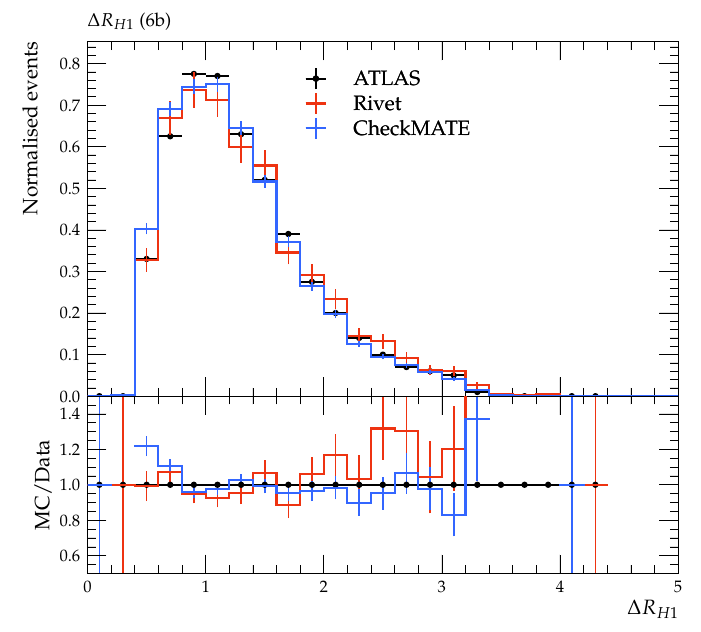}
        \caption{$\Delta R_{H1}$}
        \label{subfig:validationSM_DeltaRH1}
    \end{subfigure}
    \begin{subfigure}{0.27\textwidth}
        \centering
        \includegraphics[width=\linewidth]{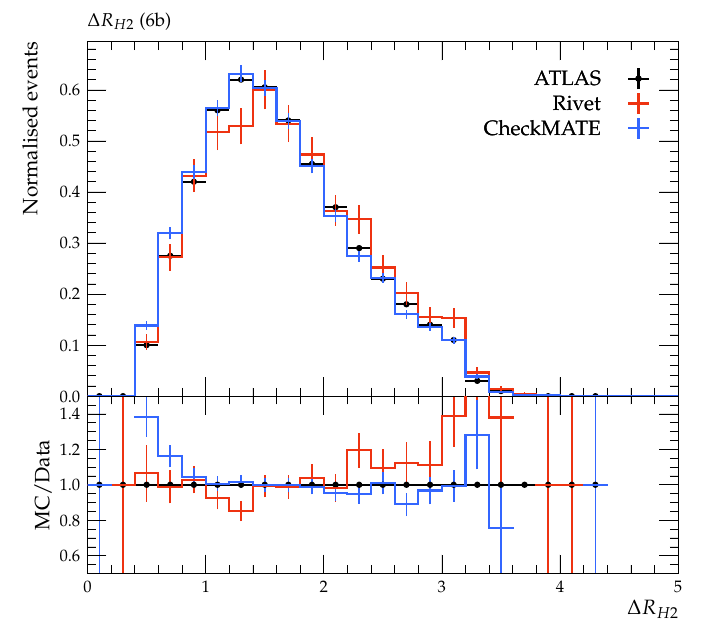}
        \caption{$\Delta R_{H2}$}
        \label{subfig:validationSM_DeltaRH2}
    \end{subfigure}
    
    \vspace{1em}
    
    % --- Second row (3 images) ---
    \begin{subfigure}{0.27\textwidth}
        \centering
        \includegraphics[width=\linewidth]{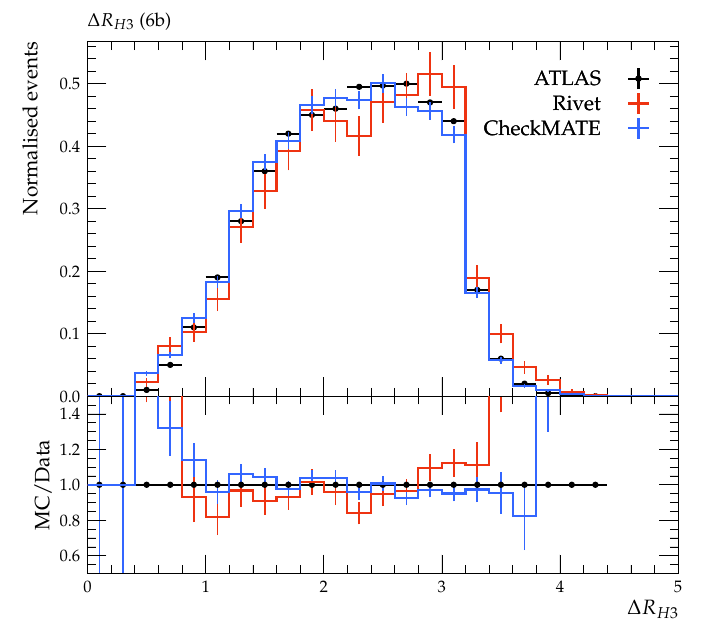}
        \caption{$\Delta R_{H3}$}
        \label{subfig:validationSM_DeltaRH3}
    \end{subfigure}
    \begin{subfigure}{0.27\textwidth}
        \centering
        \includegraphics[width=\linewidth]{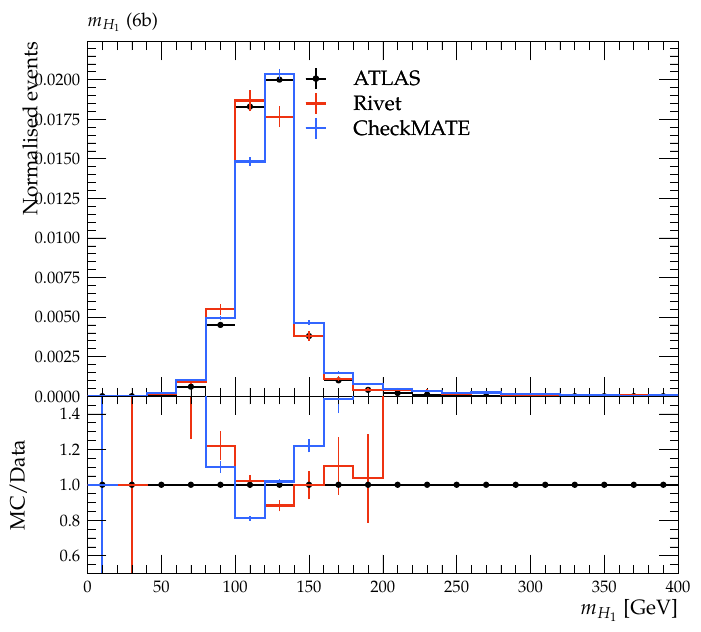}
        \caption{$m_{H1}$}
        \label{subfig:validationSM_mH1}
    \end{subfigure}
    \begin{subfigure}{0.27\textwidth}
        \centering
        \includegraphics[width=\linewidth]{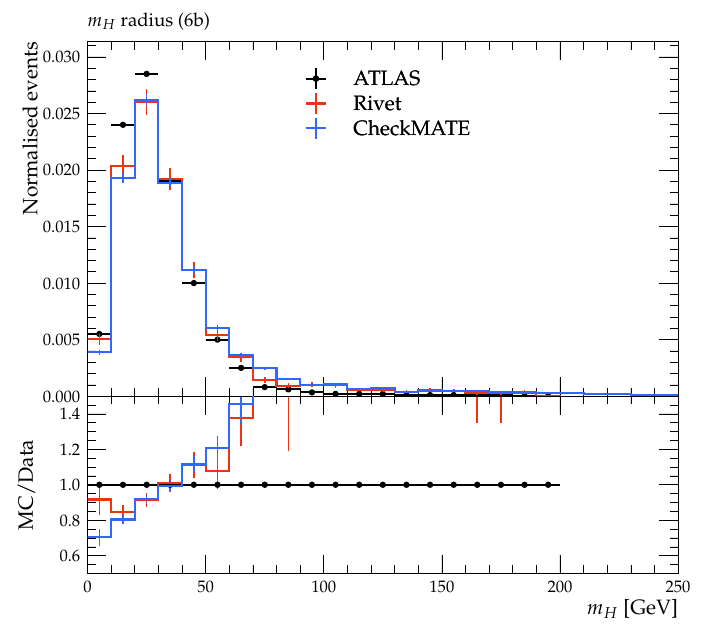}
        \caption{$m_{H}$-radius}
        \label{subfig:validationSM_mH_radius}
    \end{subfigure}
    
    \vspace{1em}
    
    % --- Third row (3 images) ---

    \begin{subfigure}{0.27\textwidth}
        \centering
        \includegraphics[width=\linewidth]{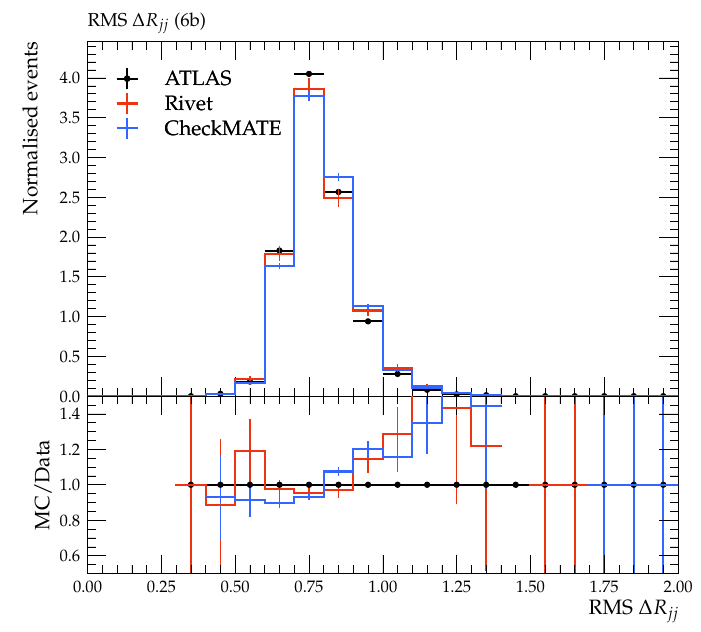}
        \caption{RMS ($\Delta R_{jj}$) }
        \label{subfig:validationSM_RMSRjj}
    \end{subfigure}
        \begin{subfigure}{0.27\textwidth}
        \centering
        \includegraphics[width=\linewidth]{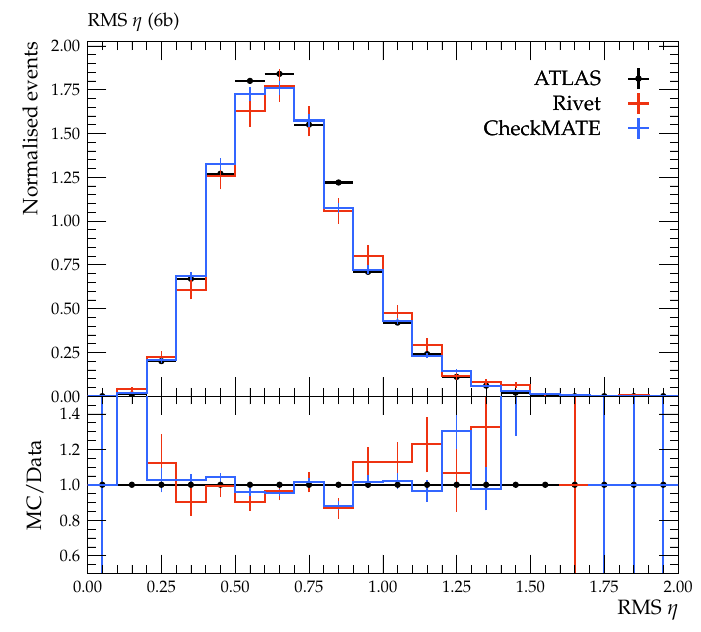}
        \caption{RMS ($\eta$)}
        \label{subfig:validationSM_RMSeta}
    \end{subfigure}
    \begin{subfigure}{0.27\textwidth}
        \centering
        \includegraphics[width=\linewidth]{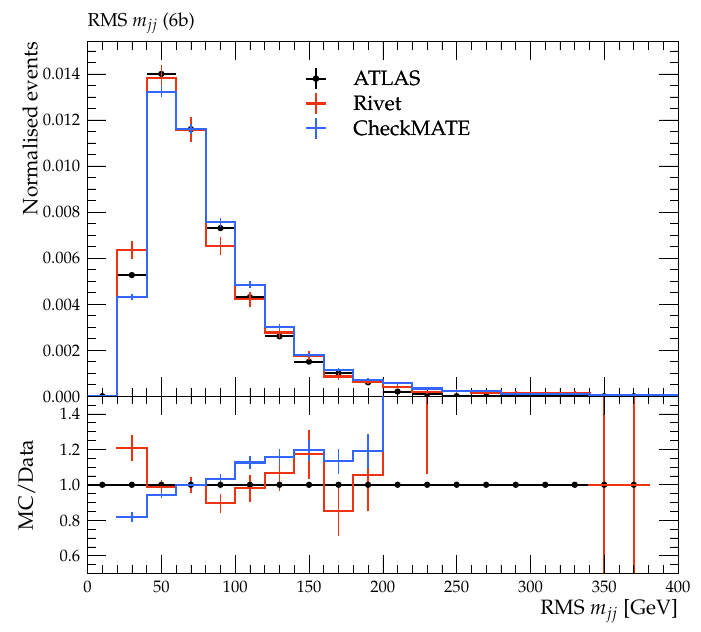}
        \caption{RMS ($m_{jj}$) }
        \label{subfig:validationSM_RMSmjj}
    \end{subfigure}
    
    \vspace{1em}
    
    % --- Last row (2 centered images) ---
    \begin{subfigure}{0.27\textwidth}
        \centering
        \includegraphics[width=\linewidth]{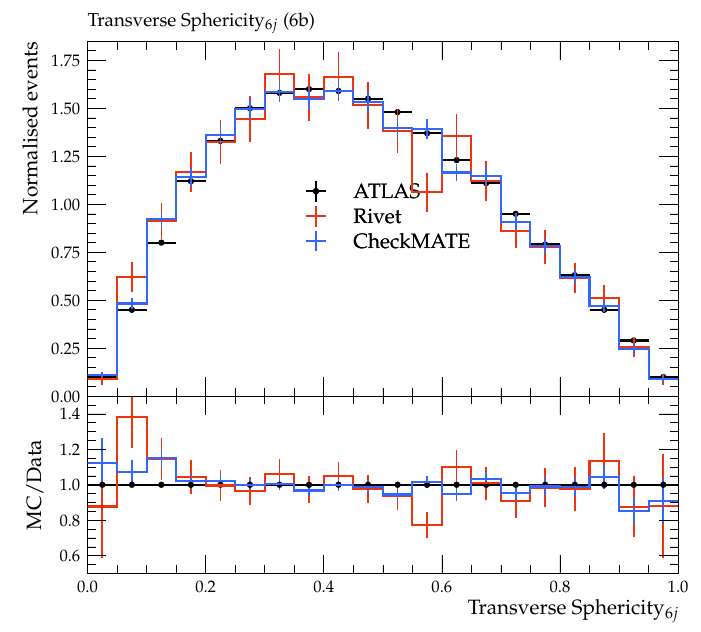}
        \caption{Transverse sphericity (6 jets)}
    \end{subfigure}\hspace{3em} % space between the two
    \begin{subfigure}{0.27\textwidth}
        \centering
        \includegraphics[width=\linewidth]{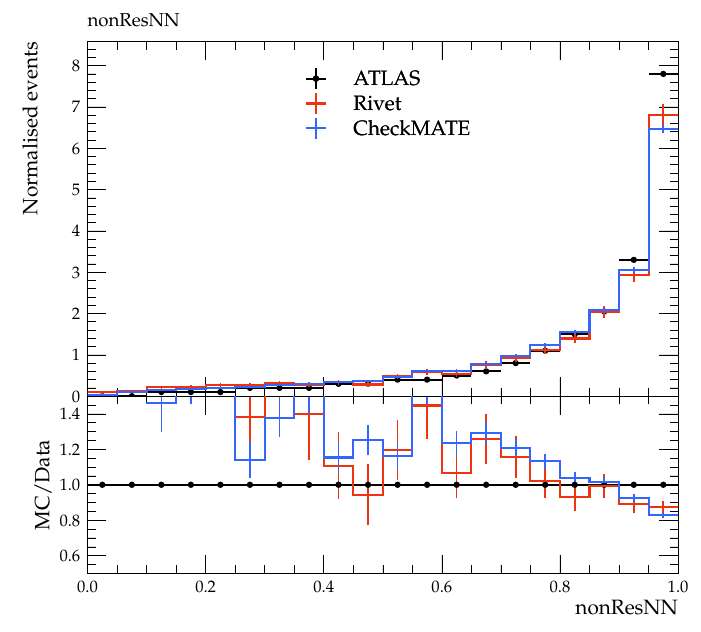}
        \caption{Non-resonant NN score}
        \label{subfig:validationSM_NN}
    \end{subfigure}
    
    \caption{The ten normalised NN inputs (a--j) and the normalised NN output (k) for the non-resonant model, for the case of SM HHH production, comparing to the results provided by ATLAS (auxiliary figures 2a--4b for the inputs and 17a for the output~\cite{HIGP-2024-32}). \rivet{} was run on 180k events and \CM{} on 200k. Both implementations show satisfactory agreement for all inputs, with differences within 20\% in all relevant bins (i.e., those with non-negligible counts). Note that the choice of binning in (k) does not accurately reflect the binning in the statistical model.} 
    \label{fig:SM_input_validation}
\end{figure}

\begin{figure}[!htbp]
    \centering
    % --- First row (3 images) ---
    \begin{subfigure}{0.27\textwidth}
        \centering
        \includegraphics[width=\linewidth]{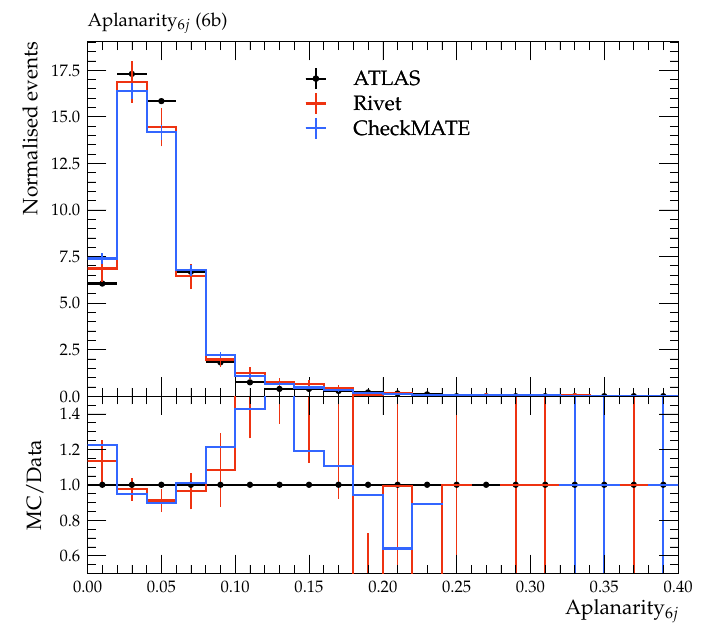}
        \caption{Aplanarity (6 jets)}
        \label{subfig:validation1500-1000_aplanarity6j}
    \end{subfigure}
    \begin{subfigure}{0.27\textwidth}
        \centering
        \includegraphics[width=\linewidth]{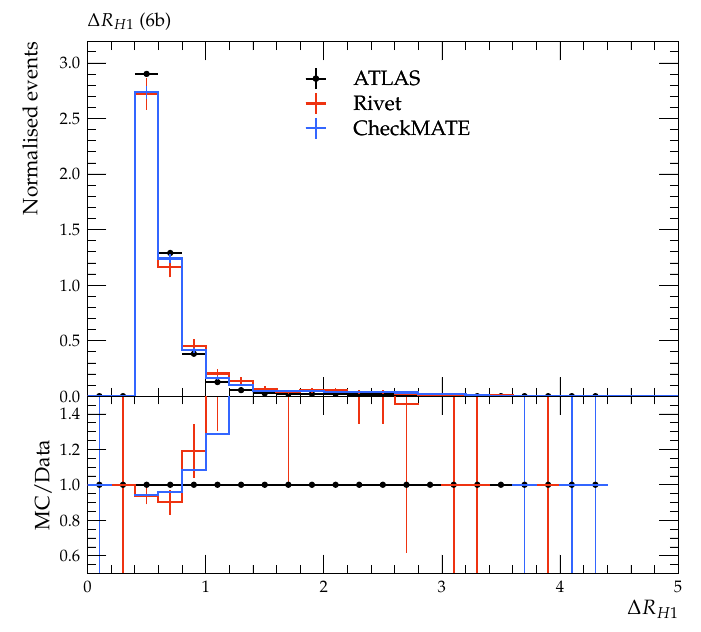}
        \caption{$\Delta R_{H1}$}
        \label{subfig:validation1500-1000_DeltaRH1}
    \end{subfigure}
    \begin{subfigure}{0.27\textwidth}
        \centering
        \includegraphics[width=\linewidth]{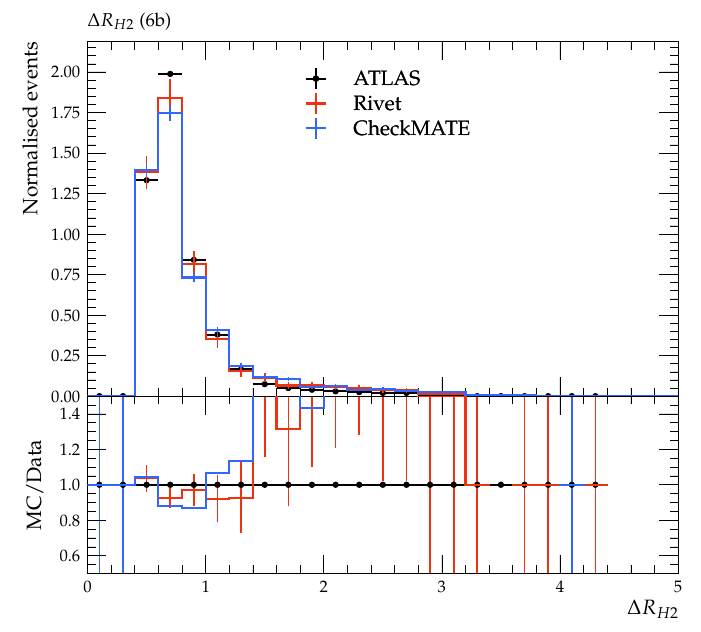}
        \caption{$\Delta R_{H2}$}
        \label{subfig:validation1500-1000_DeltaRH2}
    \end{subfigure}
    
    \vspace{1em}
    
    % --- Second row (3 images) ---
    \begin{subfigure}{0.27\textwidth}
        \centering
        \includegraphics[width=\linewidth]{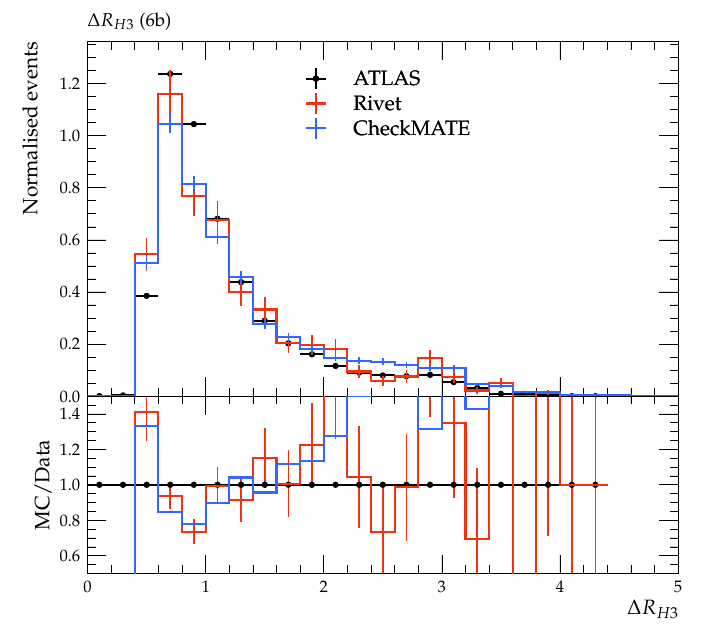}
        \caption{$\Delta R_{H3}$}
        \label{subfig:validation1500-1000_DeltaRH3}
    \end{subfigure}
    \begin{subfigure}{0.27\textwidth}
        \centering
        \includegraphics[width=\linewidth]{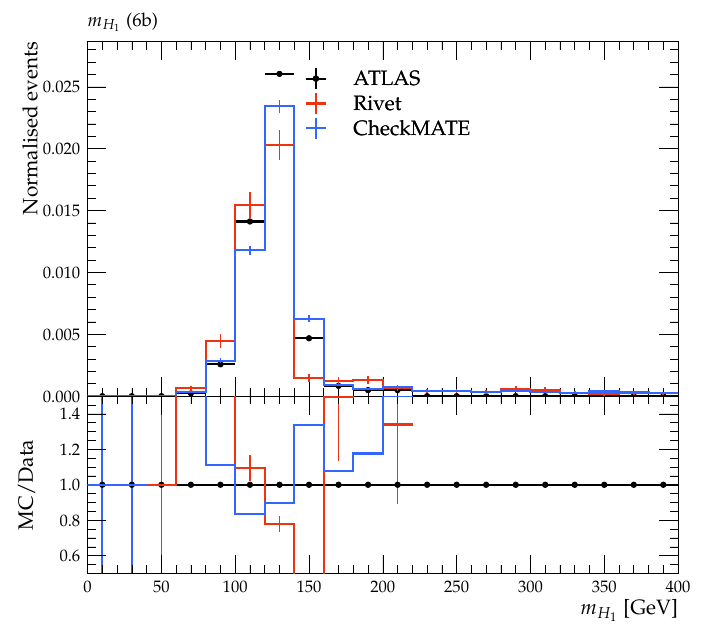}
        \caption{$m_{H1}$}
        \label{subfig:validation1500-1000_mH1}
    \end{subfigure}
    \begin{subfigure}{0.27\textwidth}
        \centering
        \includegraphics[width=\linewidth]{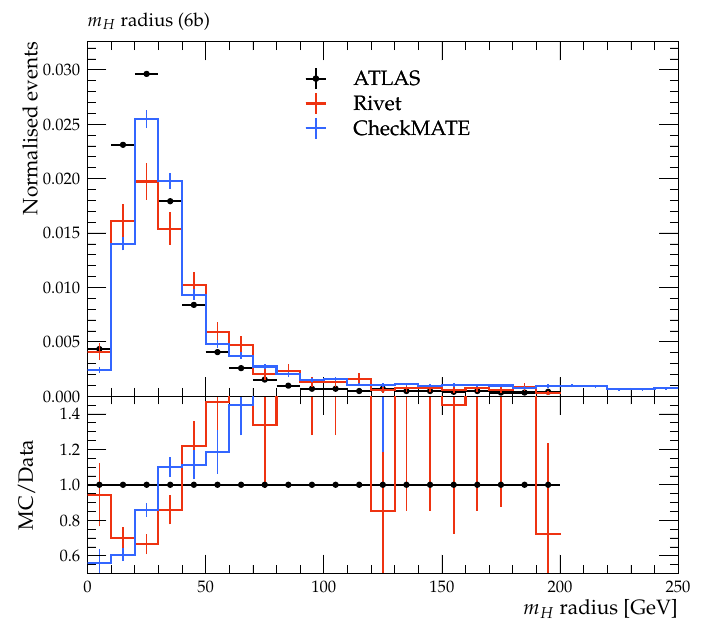}
        \caption{$m_{H}$-radius}
        \label{subfig:validation1500-1000_mHradius}
    \end{subfigure}
    
    \vspace{1em}
    
    % --- Third row (3 images) ---
    \begin{subfigure}{0.27\textwidth}
        \centering
        \includegraphics[width=\linewidth]{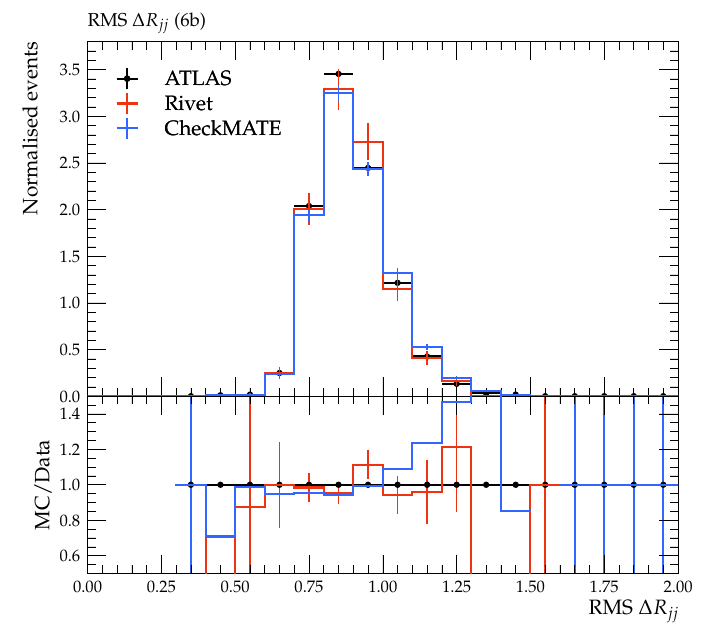}
        \caption{RMS ($\Delta R_{jj}$) }
        \label{subfig:validation1500-1000_RMSDeltaRjj}
    \end{subfigure}
    \begin{subfigure}{0.27\textwidth}
        \centering
        \includegraphics[width=\linewidth]{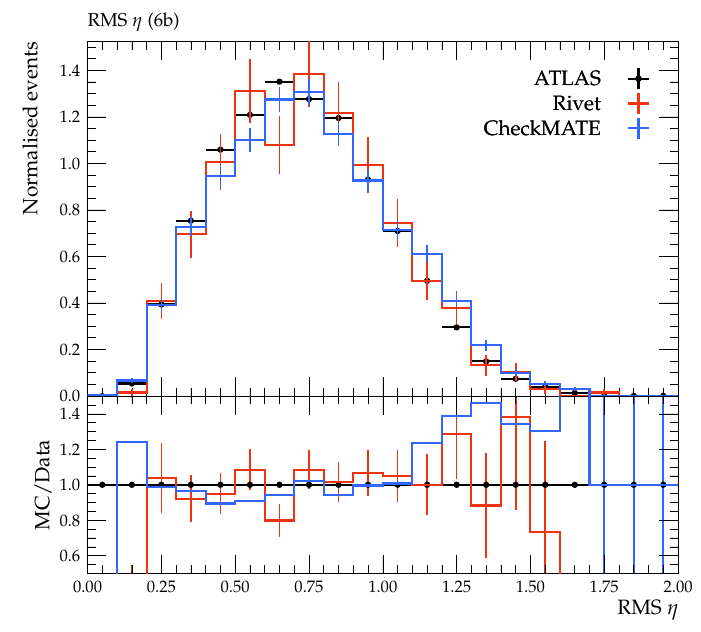}
        \caption{RMS ($\eta$) }
        \label{subfig:validation1500-1000_RMSeta}
    \end{subfigure}
    \begin{subfigure}{0.27\textwidth}
        \centering
        \includegraphics[width=\linewidth]{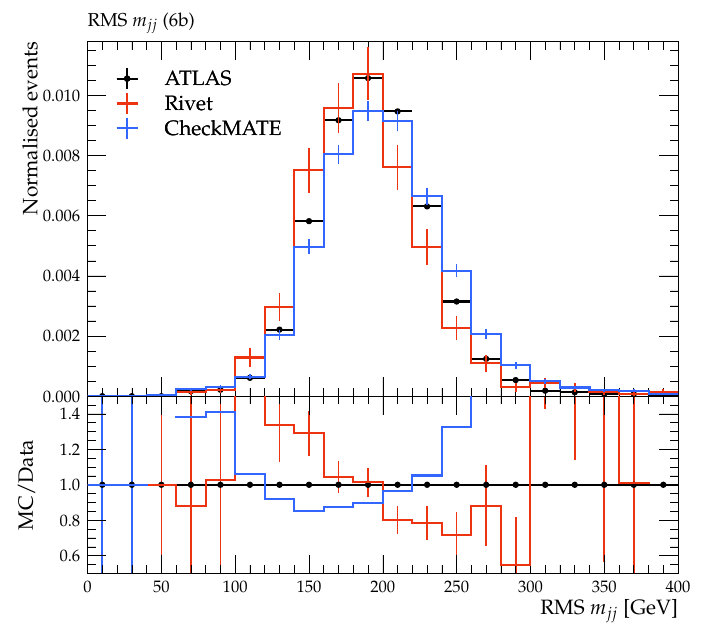}
        \caption{RMS ($m_{jj}$) }
        \label{subfig:validation1500-1000_RMSmjj}
    \end{subfigure}
    
    \vspace{1em}
    
    % --- Last row (2 centered images) ---
    \begin{subfigure}{0.27\textwidth}
        \centering
        \includegraphics[width=\linewidth]{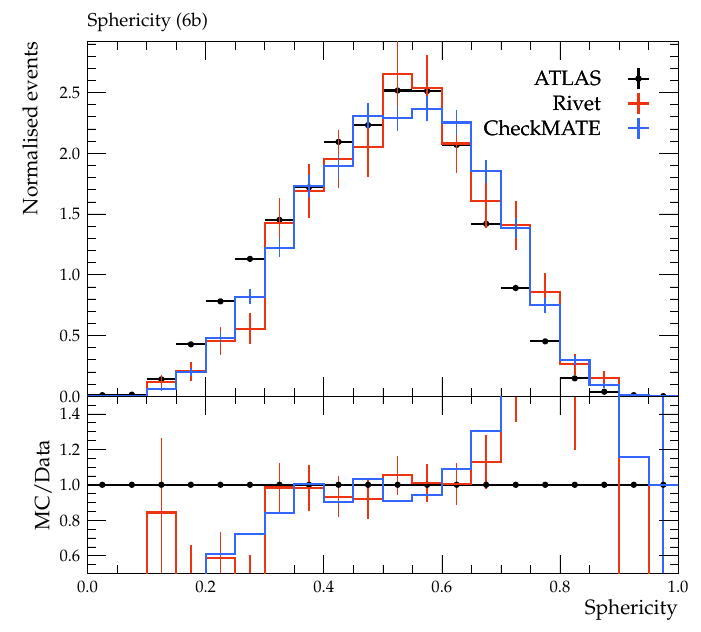}
        \caption{Sphericity (all jets)}
    \end{subfigure}\hspace{3em} % space between the two
    \begin{subfigure}{0.27\textwidth}
        \centering
        \includegraphics[width=\linewidth]{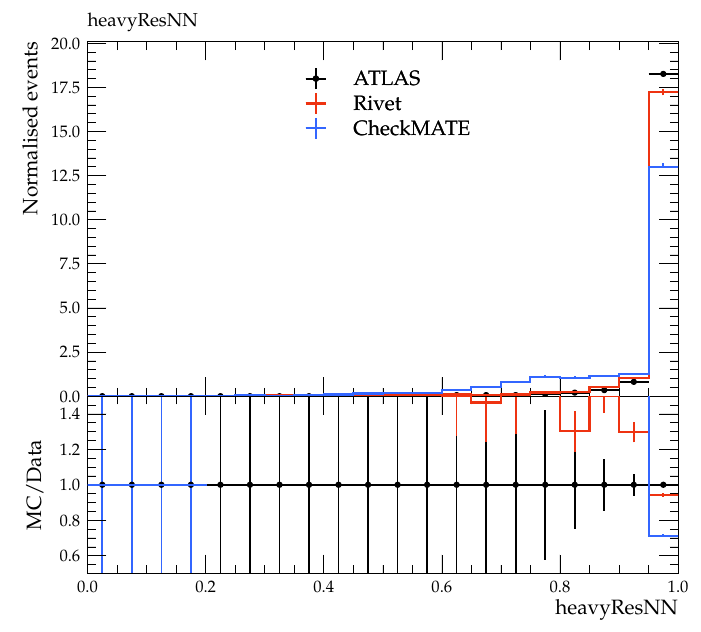}
        \caption{Heavy res. NN score}
        \label{subfig:validation1500-1000_NN}
    \end{subfigure}
    
    \caption{The ten normalised NN inputs (a--j) and the normalised NN output (k) for the heavy-resonant model, for the (1000, 1500) mass point (narrow width), comparing to the results provided by ATLAS (auxiliary figures 8a--10b for the inputs and 17c for the output~\cite{HIGP-2024-32}). \rivet{} was run on 40k events and \CM{} on 200k. Both implementations show satisfactory agreement for all inputs, with differences within 20\% in all relevant bins (i.e., those with non-negligible counts), although notably this extreme mass point has the worst NN score performance of all points.}
    \label{fig:TRSM_1500_1000_validation}
\end{figure}

In this section, we compare the set of input features from ATLAS (from auxiliary Figures~2 to~13 in Ref.~\cite{HIGP-2024-32}) against the equivalents from \rivet{} and \CM{} for all three neural nets used: the resonant (Figure~\ref{fig:TRSM_450_275_validation}); non-resonant (Figure~\ref{fig:SM_input_validation}); and heavy-resonant (Figure~\ref{fig:TRSM_1500_1000_validation}). The agreement of both \CM{} and \rivet{} with the ATLAS signals is consistently very good, especially in most of the distributions.

%%%%%%%%%%%%%%%%%%%%%%%%%%%%%%%%%%%%%%%%%%%%%%%%%%%%%%%%%%%%%%
% 450,275 plots

%%%%%%%%%%%%%%%%%%%%%%%%%%%%%%%%%%%%%%%%%%%%%%%%%%%%%%%%%%%%%%
% SM plots

%%%%%%%%%%%%%%%%%%%%%%%%%%%%%%%%%%%%%%%%%%%%%%%%%%%%%%%%%%%%%%
% 1500,1000 plots

During the process of recasting, two main discrepancies were found between descriptions in the published paper (and associated auxiliary data, including the \textsc{SimpleAnalysis} code) and what appears was actually carried out during the original study. We should emphasise that this is not a criticism of the analysis team: indeed, these issues were only discoverable due to the wealth of reinterpretation material provided, and were only fixable due to the team's determination to follow up on our questions. However, for the sake of completeness and to aid other reinterpretation efforts going forward, we will briefly describe the issues encountered.

The most significant discrepancy related to the aplanarity, sphericity, and transverse sphericity of the leading six jets (defined in Table~\ref{tab:inputDNN}): variables which were used as inputs to the resonant and non-resonant DNNs.\footnote{Both DNNs used the aplanarity, whereas only the resonant model used the sphericity and only the non-resonant model used the transverse sphericity.} The original analysis in fact calculated these observables after the leading 6-jets had been boosted into the rest-frame of the triple-Higgs system. Identifying this issue led to a significant improvement in the validation of both the DNN inputs and the final DNN score, as shown in Figure~\ref{fig:BoostImpact}.

A similar issue affected the $\mathrm{RMS}(\eta)$ variable. Table~1 of the ATLAS paper~\cite{ATLAS:2024xcs} defined it as the root-mean-square (in the ROOT sense, i.e.\ standard deviation) of the pseudorapidity of the three Higgs-candidates. In fact, it is the root-mean-square (still in the ROOT sense) of the six-leading $b$-jets, \textit{before} they have been paired to form Higgs-boson candidates.

With the resolution of these issues, overall the agreement across almost all the normalised inputs and outputs is very good for all three neural networks: the resonant (Figure~\ref{fig:TRSM_450_275_validation}); non-resonant (Figure~\ref{fig:SM_input_validation}); and heavy-resonant (Figure~\ref{fig:TRSM_1500_1000_validation}). In particular, the strong performance in reproducing the outputs (Figures~\ref{subfig:validation450-275_NN}, \ref{subfig:validationSM_NN}, and~\ref{subfig:validation1500-1000_NN}) suggests that the handful of distributions with slightly weaker modelling (discussed below) are of subleading importance to the neural networks.

Nevertheless, across all the sets of inputs, there were a handful of distributions that still suffer from slightly poorer agreement. We suspect $\cos\theta$ is particularly sensitive to jet reconstruction, and hence the simplified detector simulation in \CM{} (using \textsc{Delphes}) and in \rivet{} (using four-vector smearing) is likely to explain the discrepancies with ATLAS in Figure~\ref{subfig:validation450-275_CosTheta}. However, the fact that \rivet{} and \CM{} lie on either side of the ATLAS result suggests that a future, more optimised fast simulation/emulation code may be able to improve this variable should it prove important to future reinterpretation efforts, though in this case the difference in performance between \rivet{} and \CM{} did not make a significant difference to the final DNN score.

Another plot that displays slightly poorer agreement is the mass of the leading Higgs candidate, used as an input for the non-resonant and heavy-resonant models, and shown in Figures~\ref{subfig:validationSM_mH1} and ~\ref{subfig:validation1500-1000_mH1}. We believe this is largely an effect of the binning choices: based on Eq.~\eqref{equation:higgs_candidate_pairing_cost_function}, we expect this distribution to peak sharply at approximately 120~GeV, and the fact that two wide bins are divided on this value means that even a small disagreement in the central value or an asymmetry in one of the distributions can lead to much larger disagreements in the counts of the adjacent bins.

\subsection{Likelihoods}
\label{subsec:validation_likelihoods}

\begin{figure}[th]
\begin{subfigure}{0.5\textwidth}
        \centering
        \includegraphics[width=0.9\linewidth]{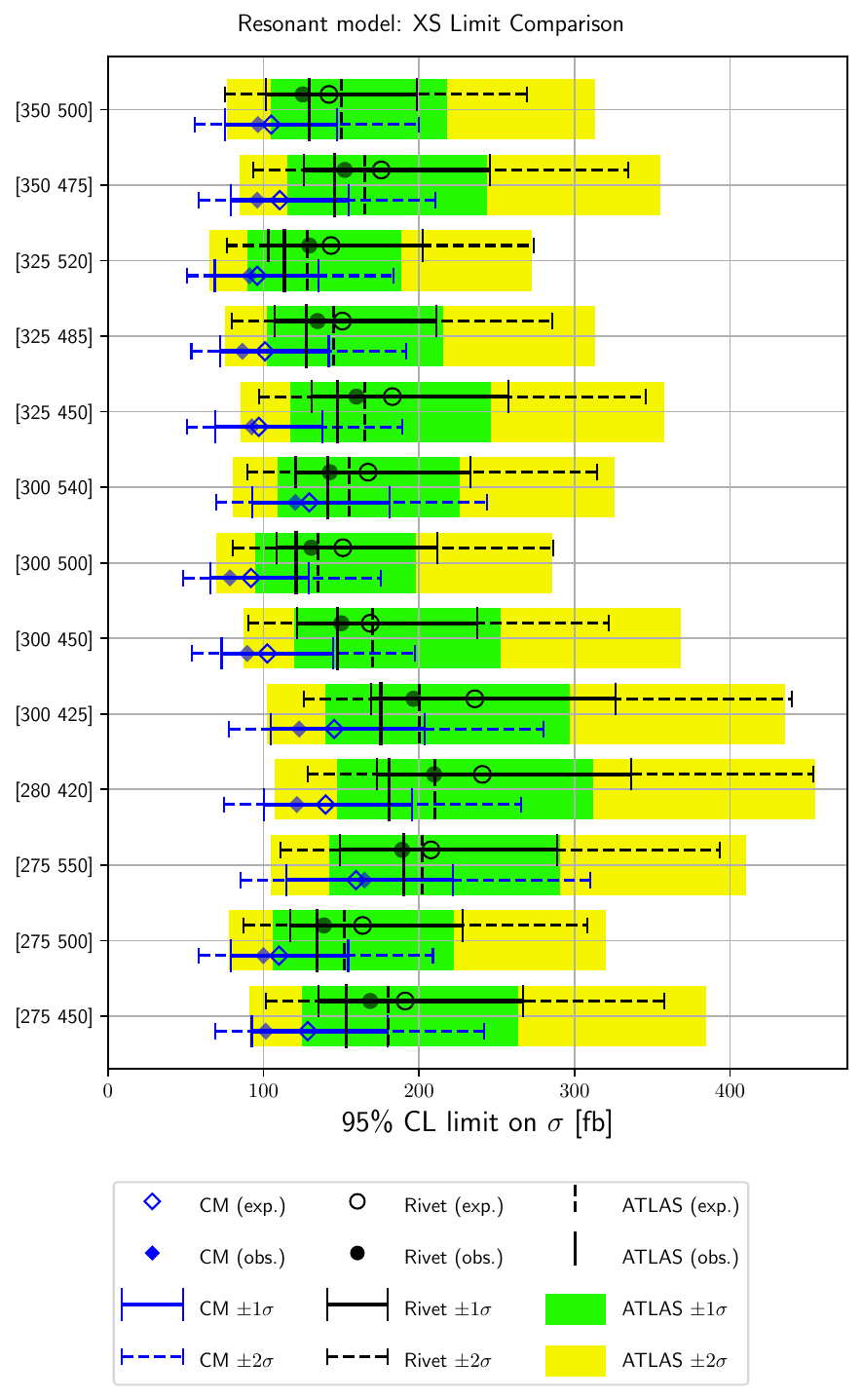}
        \caption{}
        \label{subfig:validation_res}
    \end{subfigure}
    \begin{subfigure}{0.5\textwidth}
        \centering
        \includegraphics[width=0.9\linewidth]{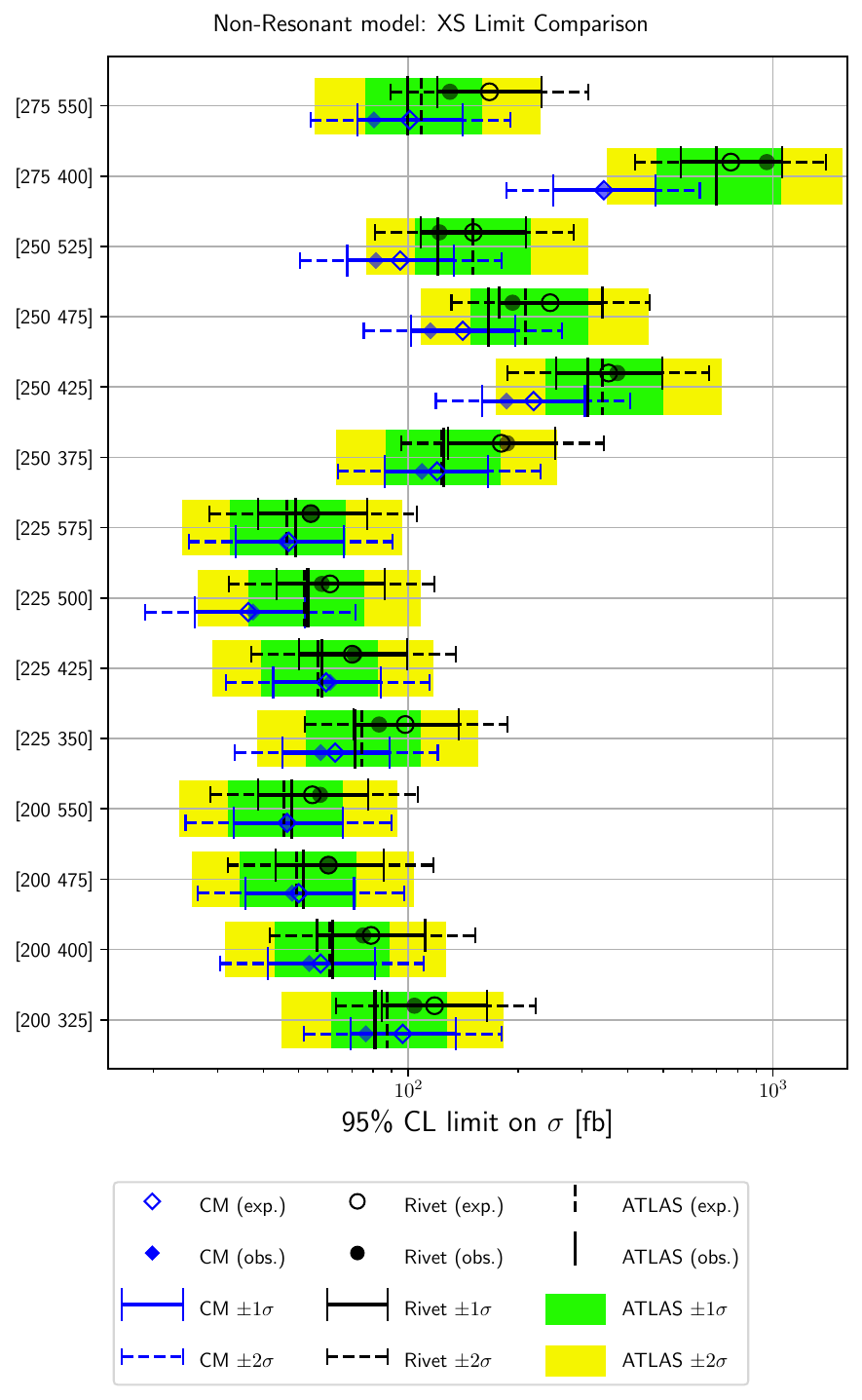}
        \caption{}
        \label{subfig:validation_nonres}
\end{subfigure}
\caption{Visualisation of the exclusion limits for the resonant model, left, and the non-resonant model, right. The ATLAS reference comes from running the HS3 files provided by ATLAS (with the parameter-of-interest (POI) range extended to ensure more stable convergence). The observed  and expected limits are similar but not identical to those published by ATLAS on HEPData. The agreement for both tools is good -- for the majority of points, both tools' central expected values lie within the ATLAS $1\sigma$ expected band. The outliers (between $1\sigma$ and $2\sigma$) are at the border or outside of the coverage of a given classifier: for example, the points $[350, 475]$ in the left panel and $[275, 400]$ in the right panel are partially non-resonant, $m_3 - m_2 = 125$~GeV, while $[275, 400]$ has $m_2 = 250$~GeV and $m_3 - m_2 > 125$~GeV. }
\label{fig:validation_res_and_nonres}
\end{figure}

\begin{figure}[pth]
\centering
\includegraphics[width=0.97\textwidth]{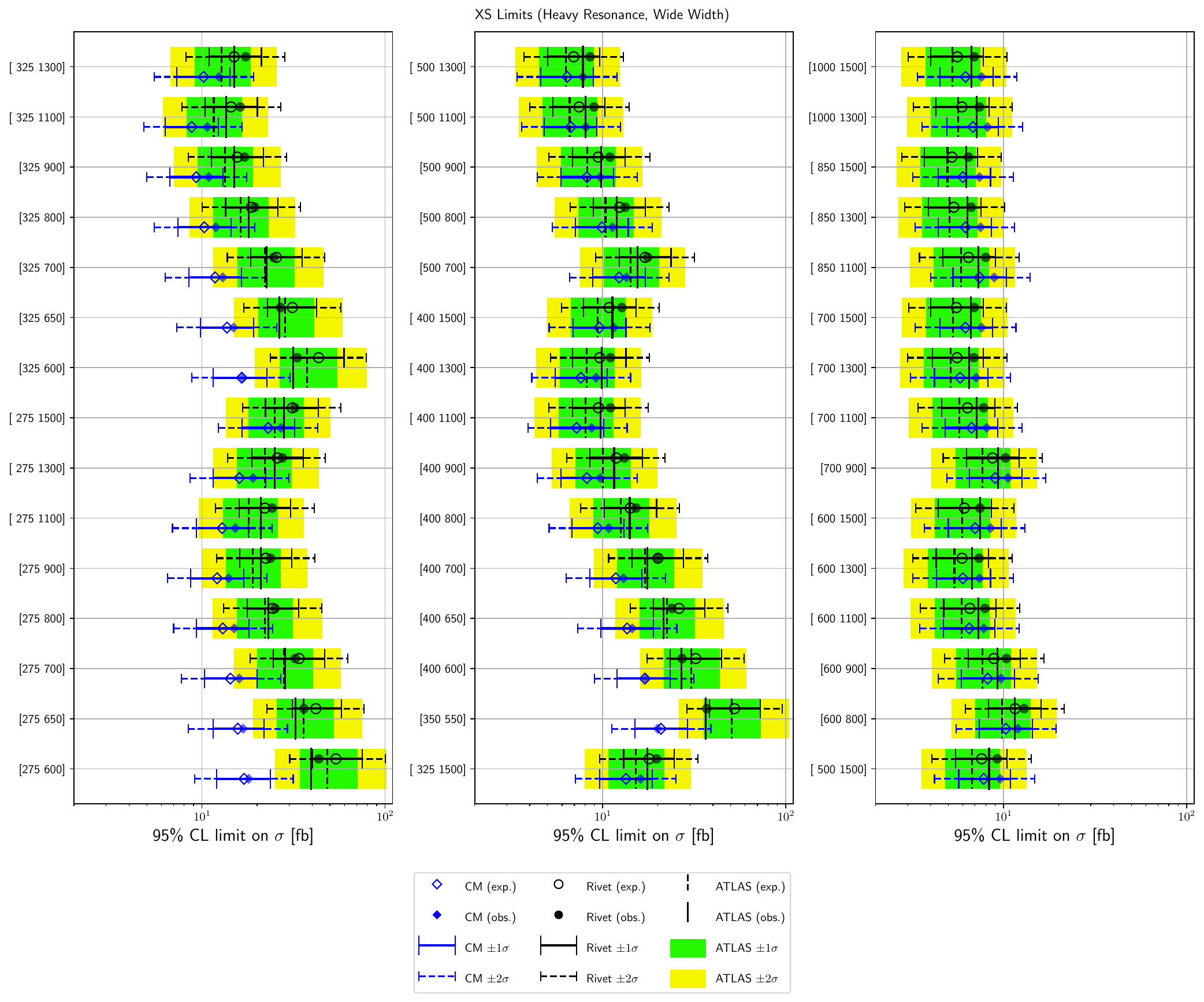}
\caption{Visualisation of the exclusion limits for the heavy-resonant model, with the wider (20\%) width. The ATLAS reference comes from running the HS3 files provided by ATLAS (with the POI range extended to ensure more stable convergence). The observed and central expected limits are similar but not identical to those published by ATLAS on HEPData (with the majority lying within the ATLAS $1\sigma$ band). The agreement for both tools is broadly good. \CM{} performs better for high-mass points and \rivet{} for low-mass points. The outliers are located near the boundary of training sample, for $m_2 \in [550, 650]$~GeV. }
\label{fig:validation_heavyres}
\end{figure}

Another method of validating the implementations in \CM{} and \rivet{} is to compare the cross-section limits obtained by ATLAS for some benchmark points with those we achieve from the reinterpretation. This was particularly straightforward in this case due to the large number of HS3 likelihood files provided by the analysis team on HEPData. By rerunning these files without changes, we could obtain ATLAS comparisons for a huge number of benchmark points (13 resonant, 14 non-resonant and 90 heavy-resonant). The limits are compared for the resonant and non-resonant models in Figures~\ref{subfig:validation_res} and~\ref{subfig:validation_nonres}, and for the heavy-resonant model in Figure~\ref{fig:validation_heavyres}.\footnote{This information is also available in tabulated form in the Zenodo record~\cite{siodmok_2026_19735119}.} Agreement is generally very good, and the central values are almost always within the error-bands of the ATLAS result. The $[-2\sigma, 2\sigma]$ range is typically slightly narrower than in the ATLAS result, as the reinterpretations do not consider all of the sources of systematic uncertainty on the signal sample, reducing the total uncertainty.

%There are some points in the heavy-resonant model that are not quite as well replicated: generally \rivet{} performs worse in the very-high mass extremes (e.g.\ the (1000, 1500)~GeV point), where it produces weaker-than-expected limits; whereas \CM{} struggles in the low-mass region, where it produces stronger-than-expected limits (e.g.\ the (275, 600)~GeV point). In both cases, these results are exacerbated by the steep gradient of the likelihood function at these points:
There are some points in the heavy-resonant model that are not quite as well replicated: in particular \CM{} struggles in the low-mass region (close to the lower limit of the training range), where it produces stronger-than-expected limits (e.g.\ the (275, 600)~GeV point). These results are further exacerbated by the steep gradient of the likelihood function at these points: in some cases a 10\% change in the observed event count changes the cross-section limit by as much as 30\%. However, for the extreme case of the parameter point (275, 600)~GeV \CM{} records about 20\% more events in the signal region and most of these go into the last bin of the DNN distribution. This increases the event count in the last bin by factor 2, and is further exacerbated in the statistical model, resulting in a stronger mean expected exclusion by factor 3. From visual inspection, there was no obvious cause for this in the distributions of any of the input features, and it appears that the response of the DNN is not linear in this edge region. %It is worth noting that both these points, (1000, 1500) and (275, 600), are in opposite corners of the probed parameter space.

\subsection{Validation summary}

In this section, we have extensively validated the \CM{} and \rivet{} implementations of the ATLAS search for $HHH$ production decaying to 6 $b$-jets~\cite{ATLAS:2024xcs}. This includes the first validation of all the input distributions to a re-interpreted neural net, enabled by the extensive auxiliary material provided by the ATLAS search team. Both the input and output distributions provided very good agreement, consistent with previous reinterpretations of detector-level searches.

The performance of statistical inference carried out based on the results of the \CM{} and \rivet{} implementations has also been extensively tested (indeed, more than 60 benchmark points may be a record), again giving results consistent with those obtained by the original analysis. This not only provides an additional confirmation that the implementations are reproducing the neural-net scores reliably, but is also particularly interesting because the sharing of likelihoods in HS3 format is relatively new, and this also provides validation for their use in reinterpretation tools and studies.

In summary, both the \rivet{} and \CM{} implementations performed very reliably, and therefore it is reasonable to use these codes to carry out new studies beyond the benchmark models provided by the original analysis, as we will now do in Sections~\ref{sec:Run2Benchmark} and~\ref{sec:hllhc-extrapolation}.

%%%%%%%%%%%%%%%%%%
%% TRSM Exclusion Plots


\afterpage{\clearpage}
\FloatBarrier
%%%%%%%%%%%%%%%%%%%%%%%%%%%%%%%%%%%%%%%%%%%%%%%%%%%%%%%%%%
\section{TRSM benchmark models at Run~2}
\label{sec:Run2Benchmark}

The obvious first use of the analysis implementations in \CM{} and \rivet{} is to study the 140 ODRB points provided in Ref.~\cite{Karkout:2024ojx}. As discussed in Section~\ref{sec:th}, because they obey perturbativity constraints, these are arguably a more ``physical'' set of models than those considered in the ATLAS paper which were used for validation in Section~\ref{subsec:validation_likelihoods}. Following a convention of Ref.~\cite{Karkout:2024ojx} we use the cross section value corrected by the NLO $K$-factor, which is about a factor 2 larger than the LO value obtained using \textsc{MadGraph}.  

\begin{figure}[tbph]
\centering
\begin{subfigure}{0.4\textwidth}
\includegraphics[width=\linewidth]{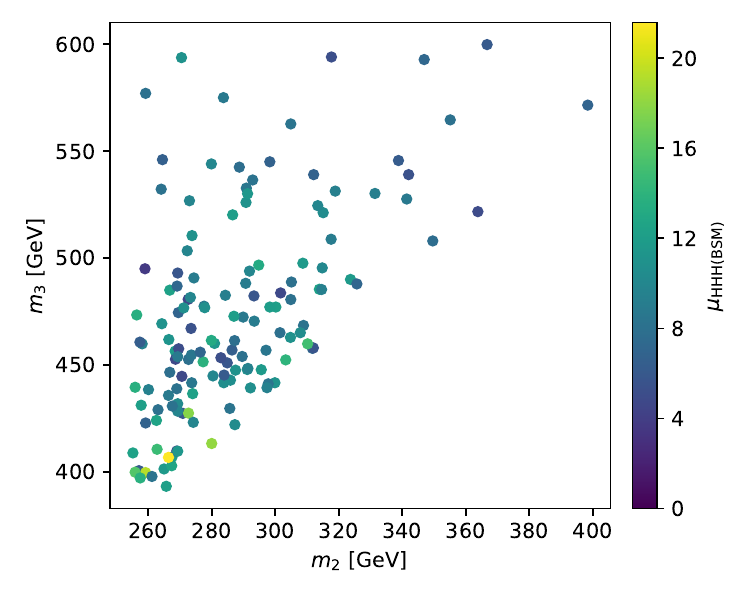}
\caption{}
\label{subfig:newbenchmark_run2_mu}
\end{subfigure}
\begin{subfigure}{0.4\textwidth}
\includegraphics[width=\linewidth]{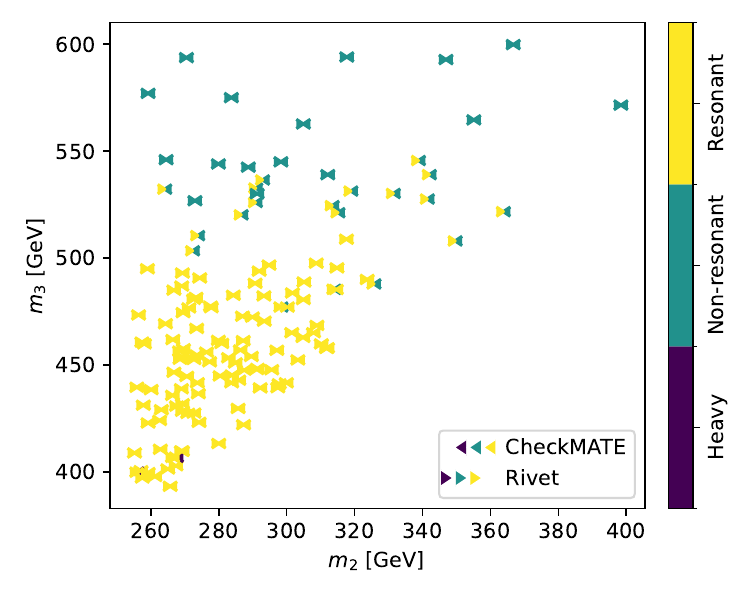}
\caption{}
\label{subfig:newbenchmark_run2_source}
\end{subfigure}

\caption{The observed limit on the signal strength $\mu$ (a) and the most sensitive neural network (b)  for the 140 benchmark points from Ref.~\cite{Karkout:2024ojx} based on the \CM{} (Figures a and b) and \rivet{} (Figure b) implementation of the ATLAS Run 2 analysis. No points are excluded (i.e.\ $\mu < 1$). While they are not displayed, the limits obtained with \rivet{} are very similar to those obtained by \CM{} in a). Figure b) shows that the strongest limit for the majority of points comes from the resonant network, with points at higher values of $m_3$ more likely to be excluded by the non-resonant network. Though it may at first appear \CM{} uses the non-resonant network a lot more, in reality where the implementations differ the difference between the two networks is usually very small and a very small shift would suffice for the strongest limit to come from the other network.} 
\label{fig:newbenchmark_run2_both}
\end{figure}

The results are shown in Figure~\ref{fig:newbenchmark_run2_both}. None of the points are excluded by the Run~2 dataset, motivating the HL-LHC extrapolation that we will carry out in Section~\ref{subsec:extrapolation_trsm_results}. The strongest observed limit on the signal strength $\mu$ is approximately 4. As shown in Figure~\ref{subfig:newbenchmark_run2_source}, for most points, the strongest limits come from the resonant DNN, with the non-resonant DNN becoming the limiting network at higher values of $m_X$. This may seem a little surprising since the non-resonant DNN was trained only on the SM signal sample, but is consistent with results from the ATLAS benchmarks used for validation in Section~\ref{subsec:validation_likelihoods}. For example, the (275, 550) mass point was studied by ATLAS with both resonant and non-resonant networks (see Figure~\ref{fig:validation_res_and_nonres}), and the 95\% CL limit on the cross-section was also a factor of two lower when using the non-resonant network, even though the point kinematically belongs to the resonant category.

It is possible that networks specifically retrained for these benchmark points could perform even better: however, we do not carry out such a procedure ourselves. Partly because this would require access to the large data sample ATLAS used for the background in training; and also because retraining on the truth-level information (even after smearing or treatment with \textsc{Delphes}) that \rivet{} and \CM{} use as input would likely produce misleadingly strong separation between signal and background, as these samples likely contain additional information that could be lost after the full detector simulation procedure. It should be noted that the process of fast detector simulation is naturally a much simpler function that would be much easier for an ML model to learn. In terms of additional truth-level information that may be present, we would be particularly concerned by correlations between the input features which may be better preserved in the MC sample. This subject is a part of the ongoing follow up studies. However we note that given the simple structure of the network in this analysis it is possible that this is not a particularly significant effect.

\FloatBarrier
%%%%%%%%%%%%%%%%%%%%%%%%%%%%%%%%%%%%%%%%%%%%%%%%%%%%%%%%
\section{Extrapolation to HL-LHC run}
\label{sec:hllhc-extrapolation}

The ATLAS experiment has published an extrapolation~\cite{ATL-PHYS-PUB-2025-003} of the results of the original 6$b$ analysis~\cite{ATLAS:2024xcs} to the HL-LHC. This only considered the SM likelihood, and was primarily concerned with the prospects for SM $HHH$ observations at the HL-LHC. It suggested that the 95\% CL limit on the SM signal strength $\mu_{\mathrm{HHH}}$ could be as low as 99 (in a scenario without systematic uncertainties), and also derived corresponding ranges for the Higgs self-coupling parameters $\kappa_3$ and $\kappa_4$.

The study did not consider any of the BP3 or ODRB TRSM models discussed above. Therefore, in this section, we describe an extrapolation of the entire analysis (i.e.\ using resonant, non-resonant and heavy-resonant neural networks) to the HL-LHC using the \CM{} and \rivet{} implementations of the analysis introduced and validated above.

\subsection{Extrapolation procedure}
\label{subsec:hllhc-extrapolation-procedure}

We closely follow the extrapolation procedure established in the ATLAS HL-LHC $HHH\to 6b$ analysis to project our sensitivity at the HL-LHC~\cite{ATL-PHYS-PUB-2025-003}. Our approach is designed to retain consistency with the published experimental methodology and enable direct comparison of results. Specifically, we repeat the following strategy:

\begin{itemize}
    \item \textbf{Luminosity scaling:} The Run~2 analysis, performed at 13~TeV with 126~fb$^{-1}$, serves as the basis for our projections. All signal and background templates will be rescaled by $\mathcal{L}_{\text{target}}/126~\text{fb}^{-1}$, where $\mathcal{L}_{\text{target}}$ is an integrated luminosity in a given scenario. We do not repeat the extrapolation at intermediate values of $\mathcal{L}_{\text{target}}$ carried out in the  ATLAS note: however, we do add an ATLAS+CMS combination luminosity of $\mathcal{L}_{\text{target}} =$ 6~ab$^{-1}$ in addition to the ATLAS-only scenario $\mathcal{L}_{\text{target}} =$ 3~ab$^{-1}$.
    
     \item \textbf{Energy and cross-section adjustments:} As in Ref.~\cite{ATL-PHYS-PUB-2025-003} the background templates (mainly QCD multijet) are further corrected by a factor of 1.18 to account for the increase in gluon-induced background at 14~TeV and Standard Model signal distributions are scaled by 1.3 according to the theoretical production cross-section ratio at 14~TeV versus 13~TeV. The BSM signal sample cross-section was obtained individually for each benchmark point, calculated as a product of $\sigma/\sigma_\textrm{SM}$ from Ref.~\cite{Karkout:2024ojx} and the SM NNLO cross section $\sigma_\textrm{SM}=0.103$ pb~\cite{deFlorian:2019app}. We checked that this value agrees within a few \% with the cross section calculated in \textsc{MadGraph} at the LO and multiplied by a $K$-factor 2.23~\cite{deFlorian:2019app}.
    
    \item \textbf{Systematic uncertainty scenarios:} We will explore a subset of the systematic uncertainty scenarios considered by the ATLAS extrapolation: a baseline HL-LHC scenario (with reduced $b$-tag/theory uncertainties compared to Run~2); an advanced scenario with reduced data-driven background uncertainties as the integrated luminosity increases; and in the most extreme case, a no-systematics scenario. It is worth noting (as mentioned in Section~\ref{subsec:implementation_cm}) that because in a recasting workflow we are not able to calculate all the systematic variations for the signal samples used by the original analysis, our ``baseline'' scenario is very similar to that we used for LHC calculations in Section~\ref{sec:validation}.
    
\end{itemize}

All extrapolated results were obtained using a binned likelihood fit implemented in HS3, based on the files provided by the original analysis using the \textit{a priori} expected limits, which should mimic as closely as possible the procedure used by ATLAS. This ensures that our projections remain robust, transparent, and in line with established experimental standards~\cite{ATL-PHYS-PUB-2025-003}.\footnote{To this end, we also provide examples of our adjusted high-luminosity HS3 files in this paper's corresponding Zenodo record~\cite{siodmok_2026_19735119}.}

\subsubsection{Advantages from access to HS3}

The publication of HS3 likelihoods alongside the analysis makes it particularly straightforward to follow the procedure used in the ATLAS extrapolation in our work. For example, samples can simply be re-scaled by multiplying all the values in a \verb|Python| list by a scale-factor. While the rescaling of systematic uncertainties is more involved and requires some additional scripting, the provision of a full likelihood means that it is possible (for what may be the first time\footnote{While previous studies have considered different methods of scaling background uncertainties when extrapolating to the HL-LHC (e.g.\ Ref.~\cite{Araz:2019otb}), they were limited by the fact that the error on the bin-by-bin background estimate provided by the experiment was listed as an individual contribution, meaning that different sources of background uncertainty could not be treated differently. The possibility of rescaling systematic uncertainties was recently included in \textsc{Spey-pyhf}~\cite{spey_pyhf_zenodo}. }) for those outside the collaboration to examine complex systematics scenarios, such as the ``reduced-background-uncertainty'' version of the baseline scenario. We hope that the further development of tools such as \textsc{pyHS3} and \textsc{Spey-HS3} may automate this procedure even more going forward, entirely removing the need to edit JSON files by hand.

\subsubsection{On the possibility of using 14~TeV signal samples}

One of the theoretical benefits of carrying out an extrapolation using new samples and a simplified workflow like \CM{} or \rivet{} (as opposed to just exploiting existing workspaces, as in the ATLAS extrapolation) is that we could, in principle, directly carry out our signal event generation at 14~TeV, forego any need to calculate scaling factors for the signal, and use this signal contribution alongside the scaled background contributions. This is potentially useful because the kinematics of the events is likely to be at least somewhat different at 14~TeV, and this may offer a greater opportunity for signal-to-background discrimination. In particular, for this analysis, events generated at a higher centre-of-mass energy are more likely to pass certain selection requirements for the SR (such as four jets with $p_T > 40$ GeV), giving a larger acceptance.

\begin{figure}[tbp]
\centering
\includegraphics[width=0.4\linewidth]{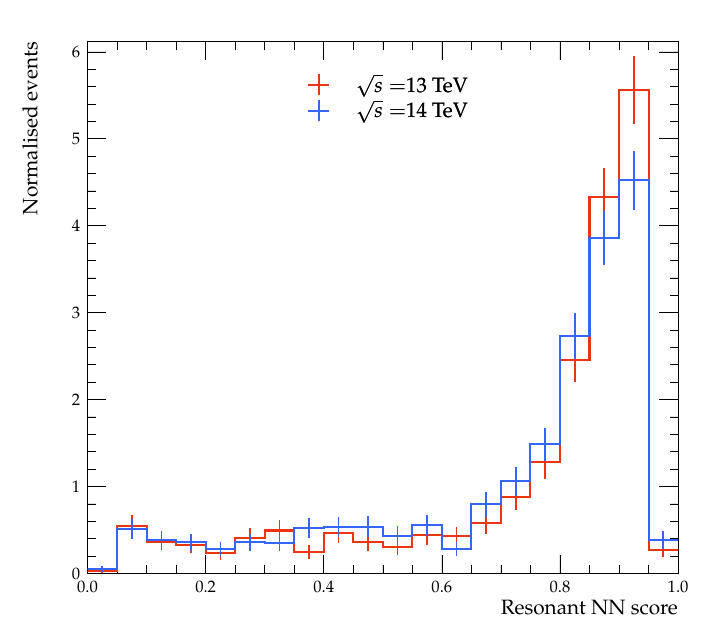}
\caption{Comparison of the resonant DNN distribution for the (287.5, 447.6) ODRB benchmark model point. Both distributions have been normalised to unity to better illustrate that the 13~TeV sample is more ``peaked".}
\label{fig:13v14_comparison_resNN}
\end{figure}

However, we found that limits obtained with 14~TeV MC were often \textit{weaker} than those found using 13~TeV MC, though there were instances where the inverse was true. This is almost certainly due to the fact that the DNNs were all trained on 13~TeV events, and so in some cases 14~TeV events in the SR may appear slightly less ``signal-like'' to the networks (as illustrated in Figure~\ref{fig:13v14_comparison_resNN} for the resonant DNN at the (287.5, 447.6) ODRB mass point). On average, this effect seems to slightly more than cancel out the increased acceptance into the SR, with the resonant DNN the most likely to be significantly affected.\footnote{This is particularly relevant because as shown for Run~2 in Figure~\ref{subfig:newbenchmark_run2_source} and for the HL-LHC in Figure~\ref{fig:hllhc-summary}, the resonant DNN provides the best limit for the majority of the new benchmark points.}

For the sake of consistency, therefore, we use the scaled 13~TeV data for our results below, although the difference in expected limits on signal strength $\mu$ is almost always within 5\%, with the two most extreme cases (one in each direction) showing differences of only 20\%. For reasons already laid out in Section~\ref{sec:Run2Benchmark} we did not consider carrying out a retraining for 14~TeV samples ourselves. However, we emphasise that re-optimising the classifier at 14~TeV for the new benchmark points has even more potential to reduce the limits that we obtain than simply re-optimising at 13~TeV as discussed in Section~\ref{sec:Run2Benchmark}.

\subsection{Standard Model Extrapolation}

\begin{table}[tb]
    \centering
    \begin{tabularx}{0.70\textwidth}{r|c|c|c}
    \toprule
    Scenario & $\mu$ (ATLAS) & $\mu$ (\rivet{}) & $\mu$ (\CM{}) \\ \midrule 
    Baseline & 195 & 185 & 201\\
    Reduced bkg.\ uncert. & 124 & 129 & 120 \\
    No systematics & 99 & 83 & 83 \\
    \midrule 
    Run 2 analysis & 750 & 780 & 745\\
\bottomrule
    \end{tabularx}
    \caption{Comparison of expected limit on $\mu_{\mathrm{HHH-SM}}$ between the ATLAS study (Table~3, Ref.~\cite{ATL-PHYS-PUB-2025-003}) and the \rivet{} and \CM{} recasting: this serves to validate our extrapolation procedure.}
    \label{tab:extrapolation-SM}
\end{table}

As a validation of the extrapolation procedure, we repeated the SM analysis carried out by ATLAS to obtain the expected limit on $\mu_{\mathrm{HHH-SM}}$, for the three scenarios discussed in the previous section. The \rivet{} and \CM{} results are compared to those obtained by ATLAS in Table~\ref{tab:extrapolation-SM}. Given the crudeness of many of the assumptions being made to carry out this extrapolation, the agreement is more than satisfactory, and so we proceed to use the same extrapolation machinery for the BSM case.

\subsection{TRSM benchmark points}
\label{subsec:extrapolation_trsm_results}

\begin{figure}[H]
\centering
\includegraphics[width=0.9\linewidth]{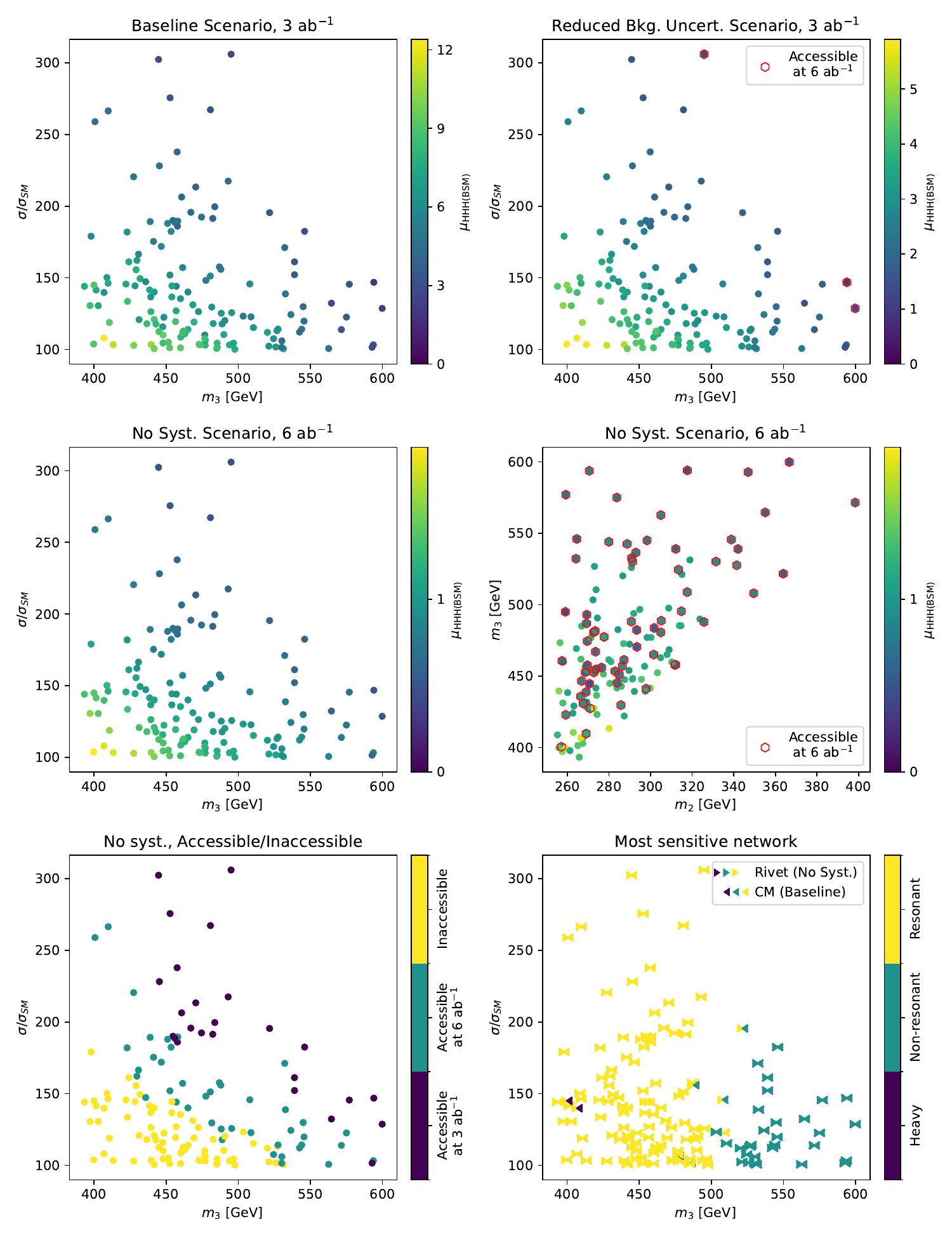}
\caption{The HL-LHC expected value of $\mu$ for the 140 ODRB models from Ref.~\cite{Karkout:2024ojx} under the three different systematic scenarios discussed in Section~\ref{subsec:hllhc-extrapolation-procedure}. The bottom-right panel also shows which neural network was used: this was relatively consistent across all systematics scenarios.}
\label{fig:hllhc-summary}
\end{figure}

The results for all 140 ODRB points under the uncertainty scenarios that we considered are shown in Figure~\ref{fig:hllhc-summary}. Note that in addition to the masses $m_2$ and $m_3$, the model has five other free parameters, so points that are closely aligned in the $(m_2, m_3)$ plane may still have different physics.

It was found that the expected exclusion tended to be stronger for points with a higher cross-section (as expected), and higher values of $m_3$ (perhaps less expected, but consistent with the Run 2 results in Section~\ref{sec:Run2Benchmark}), with only a very weak relationship between the expected exclusion and $m_2$. Therefore, the majority of the plots in Figure~\ref{fig:hllhc-summary} are displayed in the $m_3$--$\sigma/\sigma_{\mathrm{SM}}$ plane, though we include one plot (centre-right) in the $m_2$--$m_3$ projection to allow for comparison with the plots in Figure~\ref{fig:newbenchmark_run2_both} and in Ref.~\cite{Karkout:2024ojx}.

Similarly to the Run~2 case in Section~\ref{sec:Run2Benchmark}, the best limit at most points comes from the resonant DNN, with some points with $m_3$ above approximately 500~GeV being best constrained by the non-resonant DNN. The limit from the heavy-resonant network is only competitive when all three networks provide little sensitivity. This is expected since the heavy resonant DNN was optimised for larger masses than the ranges of $m_{2,3}$ in the ODRB.

No points are accessible (i.e.\ $\mu \leq 1$) in the baseline scenario. In the reduced-background-uncertainties scenario, 3 points would be accessible with a combined ATLAS+CMS data set (6~ab$^{-1}$), as highlighted in the upper-right panel of Figure~\ref{fig:hllhc-summary}. In the no-systematics scenario, 25 points would be accessible with a 3~ab$^{-1}$ data set, and almost half of the benchmark points (67 out of 140) would be accessible with the combined 6~ab$^{-1}$ data set. While the no-systematics scenario is extreme, given that we cannot account for the (potentially very significant) improvements that may come from retraining the DNN classifiers, it is nevertheless possible that this is still an under-estimate of what the HL-LHC sensitivity will be. In particular, given the decreased sensitivity at lower values of $m_3$, training a network to specifically target lower-mass points may be effective.

\section{Conclusion and outlook\label{sec:conclusion}}

In this paper we have described implementations of the ATLAS search for triple-Higgs production in the 6$b$ final state~\cite{ATLAS:2024xcs} in two different reinterpretation frameworks, \CM{} and \rivet{}. The analysis is included in the latest \CM{} release available at~Ref.~\cite{CM_github}, and it should be included in an upcoming \rivet{} release.\footnote{In the meantime, interested parties can obtain the code from the \rivet{} Merge Request~\cite{rivet_gitlab_MR}.} The implementations were compared in great detail and both thoroughly validated, including the first occasion on which all the inputs and outputs of a re-interpreted LHC analysis neural network have been individually checked. The statistical processing of the obtained yields marks the first usage of HS3 in the \CM{} framework (an external Python script was required to process the \rivet{} results), and the expected limits also provided further validation of the implementations.

The validated codes were used to study the new (optimised) doubly-resonant benchmark points suggested by Ref.~\cite{Karkout:2024ojx}. We showed that the Run~2 analysis is not sensitive to exclude any of these points, but by extrapolating our results to the HL-LHC we found that even without re-training of the neural networks, the HL-LHC will be sensitive to many of the models in at least some of the systematic scenarios considered. It is worth highlighting that the provision of the HS3-likelihoods for the Run~2 analysis by the experimental team allowed the examination of various systematic uncertainty scenarios in a level of detail hitherto impractical outside the experimental collaborations; and based on this, we provide HL-LHC limits in the case of reduced-background uncertainties or no systematic uncertainties at all, as well as the baseline case.

Stronger limits on the new benchmark points may be obtainable if the reinterpretation of the ATLAS $6b$ search could be combined with the recently published results from CMS in both the 6$b$ and 4$b$2$\gamma$ channels. In the case of both searches, we hope that the upcoming publication of the full results will be accompanied by the weights and associated metadata to allow for a thorough implementation and validation of the neural nets. Similar benefits could come from ATLAS results in other channels, or else from Run~3 results from either experiment in any channel: even though our HL-LHC extrapolation suggests that a simple Run~3 repetition of the search may not be particularly sensitive to the new benchmark points, this cannot account for signal regions which have been optimised specifically for the doubly-resonant model. There are also other physics scenarios beyond the doubly-resonant benchmark points where these reinterpretations could prove useful. In particular, it can be applied to various new physics models with extended scalar sectors, like NMSSM or N2HDM, as well as scenarios for which the BSM physics can parametrised in the EFT framework.

\section*{Acknowledgments}
The authors would like to thank Holly Pacey and Carlo Pandini for their help in understanding some of the finer details of the original ATLAS analysis and careful reading of the manuscript. We would also like to thank Jack Araz for his feedback on the paper draft and Tania Robens for her initiative to recast this study. AS, KR and TP acknowledge funding from OpenMAPP project via National Science Centre, Poland under CHIST-ERA programme (NCN 2022/04/Y/ST2/00186).
AS and TP also gratefully acknowledge Polish high-performance computing infrastructure PLGrid (HPC Center:
ACK Cyfronet AGH) for providing computer facilities and support within computational grants
PLG/2025/018203 and PLG/2026/019538. The work of AS was supported by the Priority Research Area DigiWorld under the Strategic Programme Excellence Initiative at Jagiellonian University, and was carried out within the Mark Kac Center for Complex Systems Research, Jagiellonian University, Kraków, Poland.

\section*{Summary of Zenodo resources}

This paper is accompanied by a Zenodo entry~\cite{siodmok_2026_19735119}. This entry includes:
\begin{itemize}

\item YODA files for the benchmark points used to validate the DNN inputs and outputs. We provide YODAs obtained from \rivet{} and \CM{} (\CM{} YODA was converted from the standard \CM{} output ROOT-file), as well as the ATLAS reference results from auxiliary Figures 2 to 13 of Ref.~\cite{ATLAS:2024xcs}. Notably, the ATLAS plots were not digitised on HEPData, so while our computationally-aided digitisation of the image files cannot be as accurate as the original numbers, we believe sharing these files to have the potential to be particularly helpful. These YODA files should allow for the reproduction of Figures~\ref{fig:TRSM_450_275_validation},~\ref{fig:SM_input_validation}, and~\ref{fig:TRSM_1500_1000_validation}.

\item Auxiliary figures for additional neural network validation: we provide plots for an additional resonant (325, 520) and heavy resonant point (400, 700), alongside the corresponding YODA files.

\item A summary of Run~2 limits for the BP3 benchmarks, comparing \rivet{} and \CM{} to the results from ATLAS. This should be sufficient information to reproduce Figures~\ref{fig:validation_res_and_nonres} and~\ref{fig:validation_heavyres}.

\item A summary of the limits from the ATLAS Run~2 search on the 140 benchmark points from Ref.~\cite{Karkout:2024ojx} (the ODRB points), extending the CSV file provided by that reference. This should allow for the reproduction of Figure~\ref{fig:newbenchmark_run2_both}.

\item A summary of the expected limits at the (3 ab$^{-1}$) HL-LHC for the 140 benchmark points from Ref.~\cite{Karkout:2024ojx} (the ODRB points), extending the CSV file provided by that reference. This should allow the reproduction of Figure~\ref{fig:hllhc-summary}.

\item Example \textsc{MadGraph5\_aMC@NLO 3.5.5} cards for event generation of the point (317.7, 594.0) from the ODRB set.

\item Examples in Python demonstrating how to go from a \textsc{Yoda} file produced by \rivet{} to a likelihood, both for the standard LHC case and for the more advanced case of extrapolating to the HL-LHC. This includes two example YODA files, as well as HS3 JSON files corresponding to the Run~2 likelihood (effectively the same as those provided by ATLAS, with some adjustments to signal systematics for reinterpretation) and all three systematics scenarios considered for the HL-LHC extrapolation.

\end{itemize}

%\afterpage{\clearpage}

% \input{ValidationTables/Old/HeavyNarrow}
%\newpage
\bibliographystyle{JHEP_2.18}
\bibliography{references.bib}

%\section{Appendix}
%In the following tables, we provide supplement t
%\todo[inline==True]{Do we want this as an appendix or should we just upload a supplementary dataset somewhere?}
%\input{ValidationTables/New/Resonant}
%\input{ValidationTables/New/NonResonant}
%\input{ValidationTables/New/HeavyWide}

\end{document}